\documentclass[apj]{emulateapj}

\usepackage{color}
\usepackage{CJKutf8}
\usepackage[dvipsnames]{xcolor}
\usepackage[colorlinks=true,citecolor=blue,linkcolor=blue]{hyperref}
\usepackage{dashrule}
\usepackage{subfigure}

\begin{document}

\expandafter\def\csname 0509 \endcsname{0509$-$67.5}
\expandafter\def\csname 0519 \endcsname{0519$-$69.0}
\expandafter\def\csname n103b \endcsname{N103B}
\expandafter\def\csname deml71 \endcsname{DEM\,L71}
\expandafter\def\csname 0548 \endcsname{0548$-$70.4}
\newcommand {\snr}[1]{\csname #1 \endcsname}
\newcommand {\dem}{DEM\,L\,71\ }
 
\newcommand {\bd}{Balmer-dominated\ }
\newcommand {\fline}{forbidden line }
\newcommand {\flines}{forbidden lines }

\newcommand {\xray}{X-ray\ }
\newcommand {\ha}{H$\alpha$\ }
\newcommand {\hb}{H$\beta$\ }
\newcommand {\nii}{[\ion{N}{2}]\ }
\newcommand {\hii}{\ion{H}{2}\ }
\newcommand {\oiii}{[\ion{O}{3}]\ }
\newcommand {\sii}{[\ion{S}{2}]\ }
\newcommand {\niinospace}{[\ion{N}{2}]}
\newcommand {\oiiinospace}{[\ion{O}{3}]}
\newcommand {\siinospace}{[\ion{S}{2}]}

\newcommand {\niitoha}{[\ion{N}{2}]$/$H$\alpha$\ }
\newcommand {\siitoha}{[\ion{S}{2}]$/$H$\alpha$\ }
\newcommand {\oiiitohb}{[\ion{O}{3}]$/$H$\beta$\ }
\newcommand {\oiiitoha}{[\ion{O}{3}]$/$H$\alpha$\ }

\newcommand {\kms}{km~s$^{-1}$}
\newcommand {\msun}{$M_\odot$}
\newcommand {\teff}{$T_{eff}$}

\newcommand {\s}{$\sim$\ }

\newcommand {\hst}{HST\ }
\newcommand {\chandra}{Chandra\ }
\newcommand{\gaia}{Gaia\ }

\newcommand {\cp}{\citep}
\newcommand {\ct}{\citet}
\newcommand {\clt}{\citealt}

\newcommand{\tabledashline}{ 
\hdashrule[0.5ex][x]{1.5cm}{0.1pt}{0.5mm} & \hdashrule[0.5ex][x]{2.5cm}{0.1pt}{0.5mm} & \hdashrule[0.5ex][x]{3.cm}{0.1pt}{0.5mm} & \hdashrule[0.5ex][x]{3.cm}{0.1pt}{0.5mm} & \hdashrule[0.5ex][x]{1.5cm}{0.1pt}{0.5mm} & \hdashrule[0.5ex][x]{1.8cm}{0.1pt}{0.5mm}\\}


\title{ Forbidden Line Emission from Type Ia Supernova Remnants\\ Containing Balmer-Dominated Shells}
 
\author{Chuan-Jui Li \begin{CJK}{UTF8}{bsmi}(李傳睿)\end{CJK}\altaffilmark{1}, 
You-Hua Chu \begin{CJK}{UTF8}{bsmi}(朱有花)\end{CJK}\altaffilmark{1,2}, John C. Raymond\altaffilmark{3},\\
Bruno Leibundgut\altaffilmark{4}, Ivo R. Seitenzahl\altaffilmark{5}, Giovanni Morlino\altaffilmark{6}
}
 
\affil{$^1$ Institute of Astronomy and Astrophysics, Academia Sinica, No.\ 1, Sec. 4, Roosevelt Rd., Taipei 10617, Taiwan\ 
\\ cjli@asiaa.sinica.edu.tw, yhchu@asiaa.sinica.edu.tw\\
$2$ Department of Astronomy, University of Illinois at Urbana-Champaign, 1002 West Green Street, \\
Urbana,IL 61801, U.S.A. \\
$^3$ Harvard-Smithsonian Center for Astrophysics, 60 Garden Street, Cambridge, MA, 02138, USA\\
$^4$ European Southern Observatory, Karl-Schwarzschild-Straße 2, 85748 Garching bei München, Germany\\
$^5$ School of Science, University of New South Wales, Australian Defence Force Academy,\\
Canberra, ACT 2600, Australia\\
$^6$ INAF/Osservatorio Astrofico di Arcetri, L.go E. Fermi 5, 50125 Firenze, Italy
}

%

\begin{abstract}
Balmer-dominated shells in supernova remnants (SNRs) are produced by 
collisionless shocks advancing into a partially neutral medium,
and are most frequently associated with Type Ia supernovae.
We have analyzed Hubble Space Telescope (HST) images and 
VLT/MUSE or AAT/WiFeS observations of five  Type Ia SNRs containing
Balmer-dominated shells in the LMC: 
\snr{0509}, \snr{0519}, N103B, DEM\,L71, and \snr{0548}.
Contrary to expectations, we find bright forbidden line 
emission from small dense knots embedded in four of these SNRs.
The electron densities in some knots are higher than 10$^4$ 
cm$^{-3}$.  The size and density of these knots are not
characteristic for interstellar medium (ISM) -- they most likely
originate from a circumstellar medium (CSM) ejected by the SN progenitor.
Physical property variations of dense knots in the SNRs
appear to reflect an evolutionary effect.  The recombination timescales
for high densities are short, and HST images of N103B taken 3.5 yr 
apart already show brightness changes in some knots.  VLT/MUSE 
observations detect [\ion{Fe}{14}] line emission from reverse shocks
into SN ejecta as well as forward shocks into the dense knots.
Faint [\ion{O}{3}] line emission is also detected from the Balmer 
shell in \snr{0519}, N103B,
and DEM\,L71. 
We exclude the postshock
origin because the [\ion{O}{3}] line is narrow.  For the preshock 
origin, we considered three possibilities:
photoionization precursor, cosmic ray precursor, and neutral 
precursor.  
We conclude that the [\ion{O}{3}] emission arises from oxygen that has been photoionized by [\ion{He}{2}] $\lambda$304 photons and is then collisionally excited in a shock precursor heated mainly by cosmic rays.

\end{abstract}

\subjectheadings{ISM: supernova remnants --- Magellanic Clouds --- ISM: individual (SNR \snr{0509}, SNR \snr{0519}, SNR 0509-68.7, SNR 0505-67.9, SNR \snr{0548})}

\section{Introduction}  \label{sec:Introduction}

Supernova remnants (SNRs) are commonly identified by diffuse 
X-ray emission, nonthermal radio emission, and high [\ion{S}{2}]/\ha ratio, which are 
characteristics produced by high-velocity shocks.
However, some SNRs exhibit optical spectra that are dominated 
by hydrogen Balmer lines with no or very weak forbidden lines.
Such ``Balmer-dominated'' spectra were first observed
in the Tycho SNR \citep{Kirshner1978} and subsequently 
detected in SN 1006 \citep{Schweizer1978}, Kepler SNR 
\citep{Fesen1989}, RCW\,86 \citep{Long1990}, and Cygnus 
Loop \citep{Raymond1983} in the Galaxy, and 0509$-$67.5,
0519$-$69.0, DEM\,L71 (0505-67.9), 0548$-$70.4 \citep{Tuohy1982},
and N103B (0509-68.7) \citep{Williams2014,Li2017} 
in the Large Magellanic Cloud (LMC).  

Of the above SNRs, only the Cygnus Loop originates from a 
core-collapse supernova.  Its SNR shock velocities are
low and [\ion{O}{3}] and [\ion{O}{2}] forbidden lines are seen where the shocked gas cools.  
The spectra of Cygnus Loop's Balmer 
filaments have been modeled by a $\sim$ 400 \kms\ nonradiative shock \citep{Medina2014}. 
The other SNRs containing Balmer-dominated shells are all of Type Ia
and are young or relatively young.
The weakness or absence of forbidden lines and the observed
narrow core and broad wings of their Balmer line profiles 
\citep{Smith1991} can be explained by collisionless shocks
advancing into a partially neutral medium \citep{Chevalier1980}.  
The interstellar neutral H atoms enter the shock front and
can be collisionally excited and emit Balmer lines, forming 
the narrow core, while the postshock thermalized interstellar
protons can go through charge exchange with the neutrals and 
emit Balmer lines, forming the broad wings \citep{Heng2010}.

For some of the Type Ia SNRs with Balmer-dominated shells/filaments, 
forbidden line
emission is detected at a significant level.  In the cases
of Kepler \citep{Blair1991, Sankrit2008} and N103B \citep{Li2017}, bright
forbidden lines are detected from dense knots that represent
a circumstellar medium (CSM) ejected by the progenitor before 
the SN explosion, indicating that the progenitor white dwarf
must have accreted material from a normal star companion. 

Faint forbidden lines associated with the forward and reverse shocks
in Type Ia SNRs with Balmer-dominated shells have also been reported.
For example, [\ion{O}{1}] $\lambda$6300 line emission associated 
with the Balmer shell has been reported in N103B and suggested 
to be excited in the cosmic ray precursors of the Balmer-dominated 
forward shocks \citep{Ghavamian2017}.  
For SNRs with a dense CSM component, such as N103B, the forward 
radiative or partially radiative shocks into the CSM at shock speeds 
of 350--450 km s$^{-1}$ can excite coronal [\ion{Fe}{14}] $\lambda$5303 
and other coronal lines, as seen in N103B \citep{Ghavamian2017}.
Similarly, reverse shocks driven into the SN
ejecta at 350--450 km s$^{-1}$ speed can also excite coronal [\ion{Fe}{14}] line emission,
as reported in SNRs 0509--67.5, 0519--69.0, and N103B \citep{Seitenzahl2019}.
In the two largest Type Ia SNRs with Balmer-dominated shells, DEM\,L71 and 
0548$-$70.4, the presence of isolated patches of [\ion{O}{3}] 
$\lambda$5007 emission has been interpreted as an indication of
the SNR shocks becoming radiative \citep{Tuohy1982}.

\begin{figure*}
\epsscale{0.6}
\hspace{-0.5cm}
\plotone{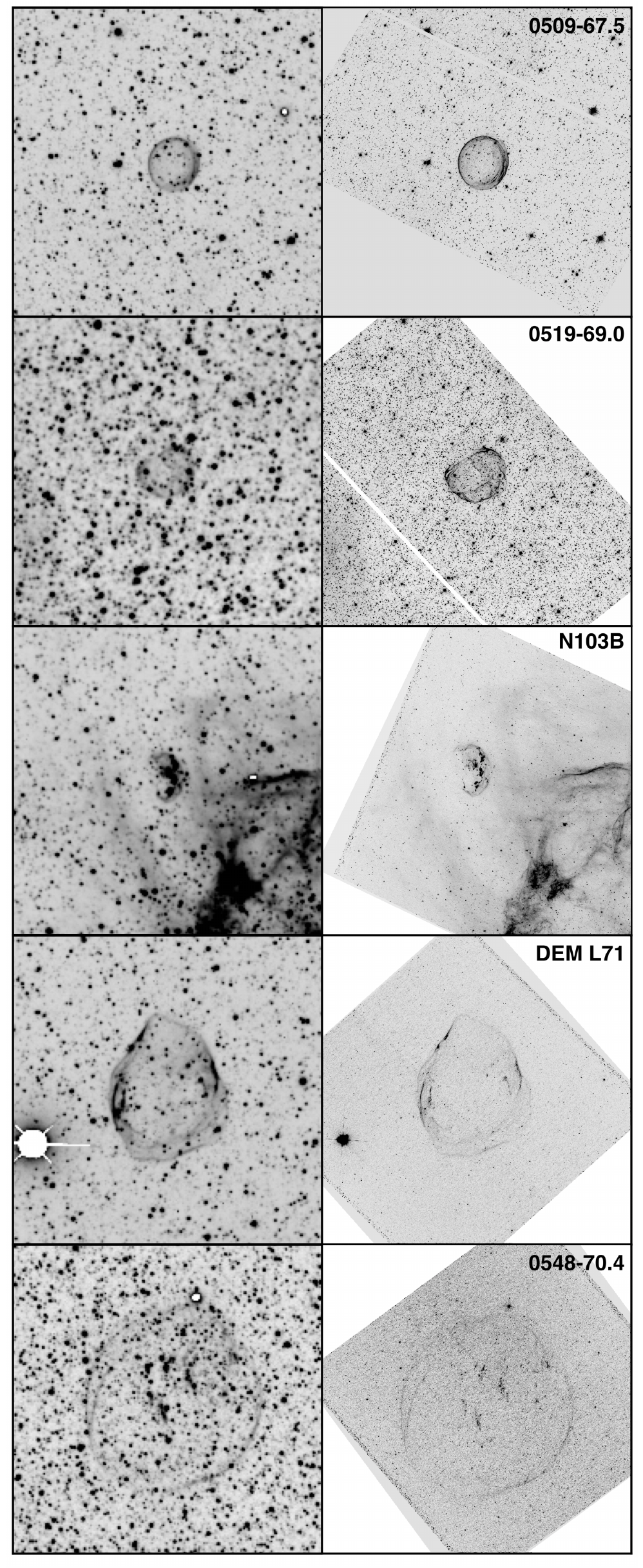}
\caption{
\ha images of the five LMC Type Ia SNRs with Balmer-dominated shells:
0509$-$67.5,  0519$-$69.0,  N103B,  
DEM L71,  and 0548$-$70.4 (from top to bottom).  Images in the
left panels were obtained with the MOSAIC II camera on the Blanco 4 m 
Telescope at Cerro Tololo Inter-American Observatory, and images in
the right panels were taken with the Hubble Space Telescope. 
The field of view of each panel is $3'\times3'$ with North at the
top and East to the left.
}
\label{figure:Ha_imgs_5SNRs_new_2}
\end{figure*}

In a program to search for surviving companions of Type Ia 
SN progenitors in the LMC \cp{Litke2017, Li2017, Li2019}, 
we have studied Hubble Space Telescope (HST) 
\ha images of the Type Ia SNRs with Balmer-dominated shells
0509$-$67.5, 0519$-$69.0, N103B, DEM\,L71, and 0548$-$70.4, 
as shown in Figure \ref{figure:Ha_imgs_5SNRs_new_2}.  
Unexpectedly, we find numerous nebular knots in DEM\,L71, 
similar to those seen in N103B and Kepler.  We have also 
used archival integral field unit type observations
of these five Type Ia SNRs to extract 
continuum-subtracted line images.
These superb imaging and spectroscopic data make it possible
to discover faint forbidden line emission and resolve dense 
knots.

This paper reports our search and analysis of forbidden
line emission from five Type Ia SNRs with Balmer-dominated 
shells in the LMC.  Section 2 describes the data used in this work, 
Section 3 introduces the method of our analysis, Section 4
reports results of individual SNRs, and Section 5 discusses
the implication of the forbidden line emission on the nature 
of the SN progenitors and the SNR shocks.  Section 6 summarizes
this study.


\section{Observations} \label{sec:Observations}  

\subsection{Hubble Space Telescope Observations}

We have obtained new HST \ha images of SNR N103B, DEM\,L71, and 
SNR 0548$-$70.4, using the UVIS channel of Wide Field Camera 3 (WFC3) 
with the F656N filter in Program 13282 (PI: Chu). The UVIS channel of 
WFC3 has a 162\arcsec\ $\times $162\arcsec\ field of view and a 
0\farcs04 pixel size.  The observations were dithered with the 
WFC3-UVIS-GAP-LINE pattern for 3 points and point spacings of 2\farcs414.  
The total exposure time for 
each SNR is 1350 s.

Archival HST \ha images of SNRs 0509$-$67.5 and 0519$-$69.0 taken 
with the Advanced Camera for Surveys (ACS) and the F658N filter are 
available in the Hubble Legacy Archive.
In addition, archival HST WFC3 \ha, [\ion{O}{3}], and [\ion{S}{2}] 
images of N103B are also available.

The HST imaging observations we use are
listed in Table \ref{table:obs}, where the filter, PI, 
program ID, date of observation, and exposure
time are given.

\subsection{VLT MUSE observations}

We have used archival Multi-Unit Spectroscopic Explorer (MUSE) 
observations obtained with the Very Large Telescope (VLT) UT4 
for SNRs 0509$-$67.5, 0519$-$69.0, N103B, and DEM\,L71.
MUSE is an integral-field unit (IFU), which  provides a spectrum 
for every position in the field of view (FOV).  For these 
observations, the FOV is 60\arcsec $\times$\,60\arcsec, large 
enough to encompass the entire SNR shell for the three small objects, 
but not DEM\,L71. The wavelength coverage, 4750 -- 9350 \AA, 
includes nebular lines such as H$\alpha$, H$\beta$, 
[\ion{O}{3}] $\lambda\lambda$4959, 5007, 
[\ion{N}{2}] $\lambda\lambda$6548, 6583, and 
[\ion{S}{2}] $\lambda\lambda$6716, 6731.
The spatial and spectral samplings are 0\farcs2 spaxel$^{-1}$ 
and 1.25 \AA\, pixel$^{-1}$, respectively.  
The archival MUSE observations are listed in 
Table \ref{table:obs} with PI, 
program ID, date of  observation, and exposure time information.

We follow the standard procedure and use the VLT MUSE data reduction pipeline \citep{Weilbacher2014} 
to carry out bias subtraction, flat fielding, and wavelength and geometrical calibrations.

\subsection{ATT WiFeS observations}


\begin{figure}
\epsscale{1.4}
\hspace{-1.5cm}
\plotone{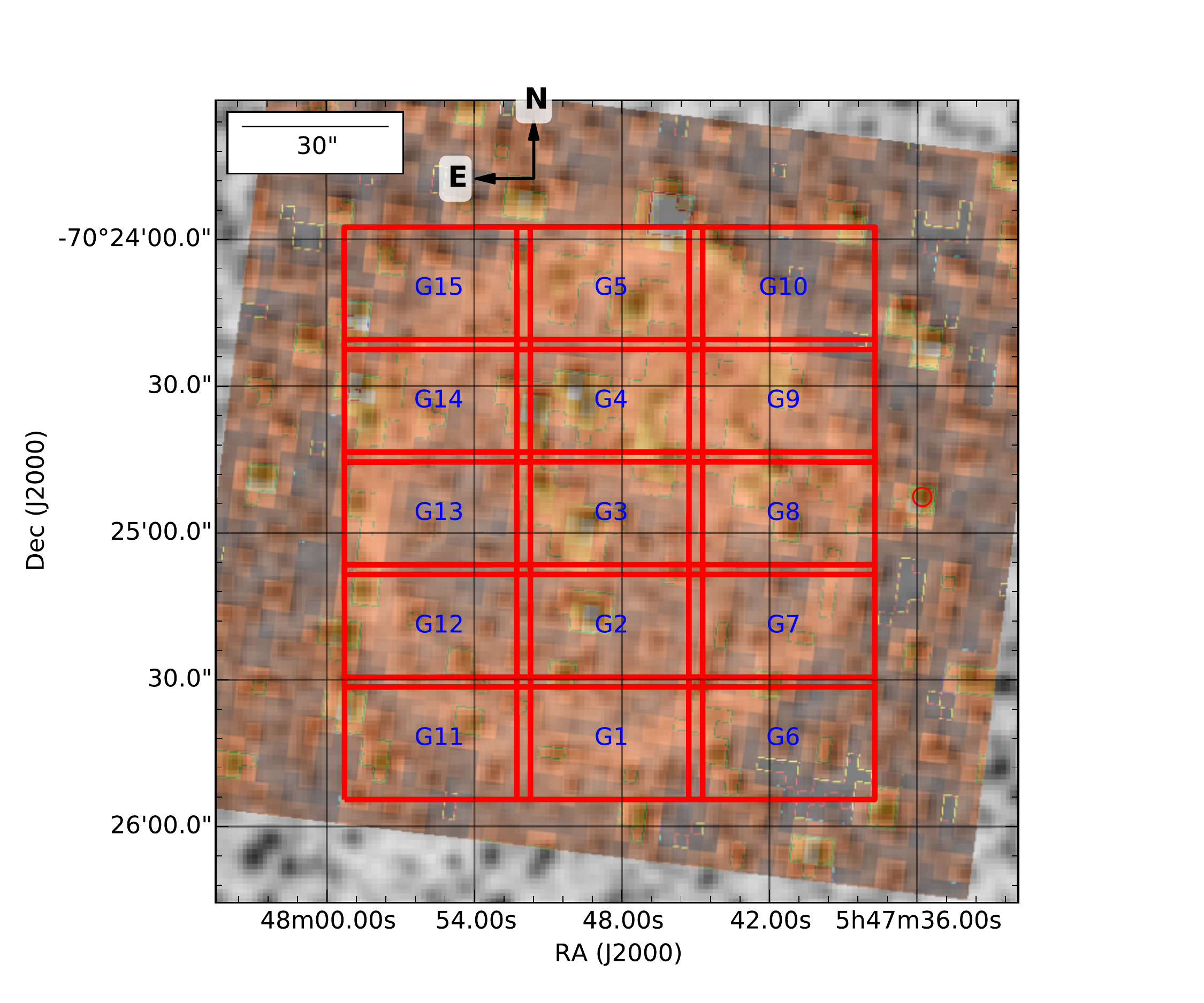}
\caption{
ATT WiFeS \ha observations of \snr{0548}. Only 
G4 and G9 data with better quality are presented 
in this paper.
}
\label{figure:SNR0548_pretty}
\end{figure}

We have used the Wide Field Integral Spectrograph (WiFeS) 
observations obtained with the Advanced Technology Telescope (ATT) for SNR \snr{0548}. 

The FOV of ATT WiFeS is 25\arcsec $\times$ 38\arcsec, much smaller than the extent of SNR \snr{0548}. Thus, observations for a grid of 3 $\times$ 5 fields were planned to map the entire SNR, as 
shown in Figure \ref{figure:SNR0548_pretty}; however, due to the weather condition, only four fields were observed. Among these observed fields, only fields G4 
and G9 have adequate quality and contain forbidden line emission to be presented 
in this study. 


The ATT WiFeS is 
a double-beam spectrograph that simultaneously covers the blue and red wavelength ranges.
For SNR \snr{0548}, the B3000 and R7000 gratings were used for the blue (3500 
-- 5700 \AA), and the red (5400 -- 7024 \AA) wavelength ranges, respectively.
The overall wavelength coverage is thus 3500 
-- 7024 \AA, which includes all major nebular lines in the optical.
The spatial sampling is 1\farcs0 $\times$ 
0\farcs5 spaxel$^{-1}$, while the spectral samplings are 0.768 \AA\ pixel$^{-1}$ for 
the blue spectral range and 0.439 \AA\ pixel$^{-1}$ for the red spectral range.
The ATT WiFeS observation of SNR \snr{0548}
is listed in Table \ref{table:obs} 
with pertinent information, such as PI, program ID, date of observation, and 
exposure time.




\begin{deluxetable*}{ccccccccc}
\tablecaption{Observations}
\tablehead{Telescope & Instrument & SNR & Filter & PI & Program ID &  Date & t$_{exp}$ (s)}
\startdata 
\hst & ACS/WFC & \snr{0509} & F658N & Hughes & 11015 & 2006 Oct 28 &  4620 \\
\hst & WFPC2 & \snr{0509} & F656N & Hughes & 11015 & 2007 Nov 07 & 14300\\
\hst & ACS/WFC & \snr{0509} & F658N & Hovey & 14733 & 2016 Nov 13 & 10662\\
\hst & ACS/WFC & \snr{0519} & F658N & Hughes & 12017 & 2011 Apr 21 & 4757 \\
\hst & WFC3/UVIS & N103B & F656N & Chu & 13282 & 2013 Jul 11 & 1350 \\ 
\hst & WFC3/UVIS & N103B & F502N & Williams & 14359 & 2017 Jan 03 & 2051 \\ 
\hst & WFC3/UVIS & N103B & F657N & Williams & 14359 & 2017 Jan 03 & 2979 \\ 
\hst & WFC3/UVIS & N103B & F673N & Williams & 14359 & 2017 Jan 03 & 1982 \\ 
\hst & WFC3/UVIS & \dem & F656N & Chu & 13282 & 2014 Mar 05 & 1350\\
\hst & WFC3/UVIS & \snr{0548} & F656N & Chu & 13282 & 2013 Sep 20 & 1350\\
VLT & MUSE & \snr{0509} & - &  Morlino & 0100.D-0151  & 2018 Jan 22 & 701 \\ 
VLT & MUSE & \snr{0519} & - &  Leibundgut & 096.D-0352 & 2016 Jan 17 & 900\\
VLT & MUSE & N103B & - &  Leibundgut & 096.D-0352  & 2015 Dec 12 & 900 \\    
VLT & MUSE & DEM L71 & -  &  Leibundgut & 096.D-0352 & 2015 Nov 16 & 900\\ 
ATT & WiFeS & \snr{0548} & - & Seitenzahl & 4150178 & 2015 Dec 14 & 1800
\enddata
\label{table:obs}
\end{deluxetable*}

\section{Method of Analysis} \label{method}  

We start our analysis with a general examination of the \hst \ha images 
of the five Type Ia SNRs with Balmer-dominated shells in the LMC.
\hst images allow us to resolve nebular features as small as 0\farcs05, 
i.e., 0.0125 pc in the LMC.  We have learned from our previous study of 
the SNR \snr{n103b} that \hst \ha images resolve two major types of 
morphological features: Balmer-dominated long filaments that delineate 
a shell structure, and forbidden-line-emitting dense knots that are 
distributed in groups in the SNR interior or along some Balmer filaments
\citep{Li2017}.
Thus, we first use the \hst \ha images to examine the shell structure and 
to search for dense nebular knots.

We then extract nebular line images from the VLT MUSE and ATT WiFeS data 
to search for nebular features that emit forbidden lines.
These IFU data cubes allow us to extract images in wavelength intervals that 
cover nebular lines as well as images in adjacent line-free wavelength intervals.  
A clean line image can be obtained by subtracting the latter from the former.  
So-extracted line images are more sensitive than ground-based  
images taken with filters because sky lines and continuum background, especially 
the stars, are both removed.  It is thus 
not surprising that we indeed have detected faint emission features that 
were not seen before.

A large number of forbidden lines of low-ionization species, such as [\ion{N}{2}], 
[\ion{O}{1}], [\ion{O}{3}], [\ion{S}{2}], [\ion{Ca}{2}], and [\ion{Ni}{2}], are
detected in the dense knots \citep{Ghavamian2017, Dopita2019},
and forbidden lines of high-ionization species, such as [\ion{Fe}{9}], [\ion{Fe}{14}], 
[\ion{Fe}{15}], and [\ion{S}{12}], are detected in nonradiative shocked ejecta
\citep{Seitenzahl2019}; however, we will focus on only the strongest 
diagnostic lines in this paper.
The line images extracted include those of H$\alpha$, [\ion{O}{3}] $\lambda$5007, 
[\ion{Fe}{14}] $\lambda$5303, [\ion{O}{1}] $\lambda$6300,
and [\ion{S}{2}] $\lambda\lambda$6716, 6731 lines.
The central wavelengths are these nebular lines red-shifted to the LMC velocity 
of $\sim$270 km s$^{-1}$.  A wavelength interval of 10 \AA\ is used 
for the H$\alpha$ line image (corresponding to $\pm$230 km s$^{-1}$) to avoid 
the [\ion{N}{2}] lines, 
8 \AA\ for [\ion{O}{3}] (corresponding to $\pm$240 km s$^{-1}$)
to accommodate 
emission at all velocities, 20 or 150 \AA\ for [\ion{Fe}{14}] (corresponding to
$\pm570$ and $\pm4250$ km s$^{-1}$, respectively) to accommodate its broad 
line profile, 15 \AA\ for [\ion{O}{1}] (corresponding to $\pm$360 km s$^{-1}$), 
and 31 \AA\ for [\ion{S}{2}] to include both $\lambda\lambda$6716, 6731 lines.  
For the line-free continuum background images,
wavelength intervals of 12--25 \AA\ are used on the two sides of each spectral line 
interval.  The average of the two spectral-background images is prorated and 
subtracted from the line image to obtain spectral-background-subtracted clean 
line image.  Throughout the rest of the paper,
only such clean line images are used and we will refer to them simply as 
``line image".

As these line images have the same FOV and image scale, they can be 
inter-compared directly to search for forbidden-line emission features. 
These \ha images can be compared with the \hst \ha images for detailed 
physical structures of  emission features. 
We note that some stars have negative values in the MUSE \ha line 
images because their spectra have the \ha line in absorption. 

Finally, we analyze the spectral properties of the forbidden-line 
emission features, measuring their line strengths relative to 
Balmer lines and using the [\ion{N}{2}] $\lambda$5755/$\lambda$6583 
diagnostic to determine electron temperatures and the 
[\ion{S}{2}] $\lambda$6716/$\lambda$6731 diagnostic to determine 
electron densities.
These physical parameters are used to assess the physical conditions 
of the forbidden-line emission features in order to determine their origin.

\section{Results of individual objects} \label{sec:results}  

The five Type Ia SNRs with Balmer-dominated shells we have studied are shown in 
Figure~\ref{figure:Ha_imgs_5SNRs_new_2}, where the same image scale is used 
for all five SNRs for easier inter-comparison.  The results of our
analyses of these five SNRs are described below.  See the Appendix for 
representative spectra of different morphological features in the SNR and 
of the background ISM, plotted in two wavelength ranges that cover
the H$\beta$ + \oiii\ lines and H$\alpha$ + \nii\ + \sii\ lines, respectively.

\begin{figure*}[htbp]
\centering
\hspace{0.6cm}
\subfigure[]{
\hspace{-2cm}
\begin{minipage}[t]{0.53\linewidth}
\centering
\includegraphics[scale=0.46]{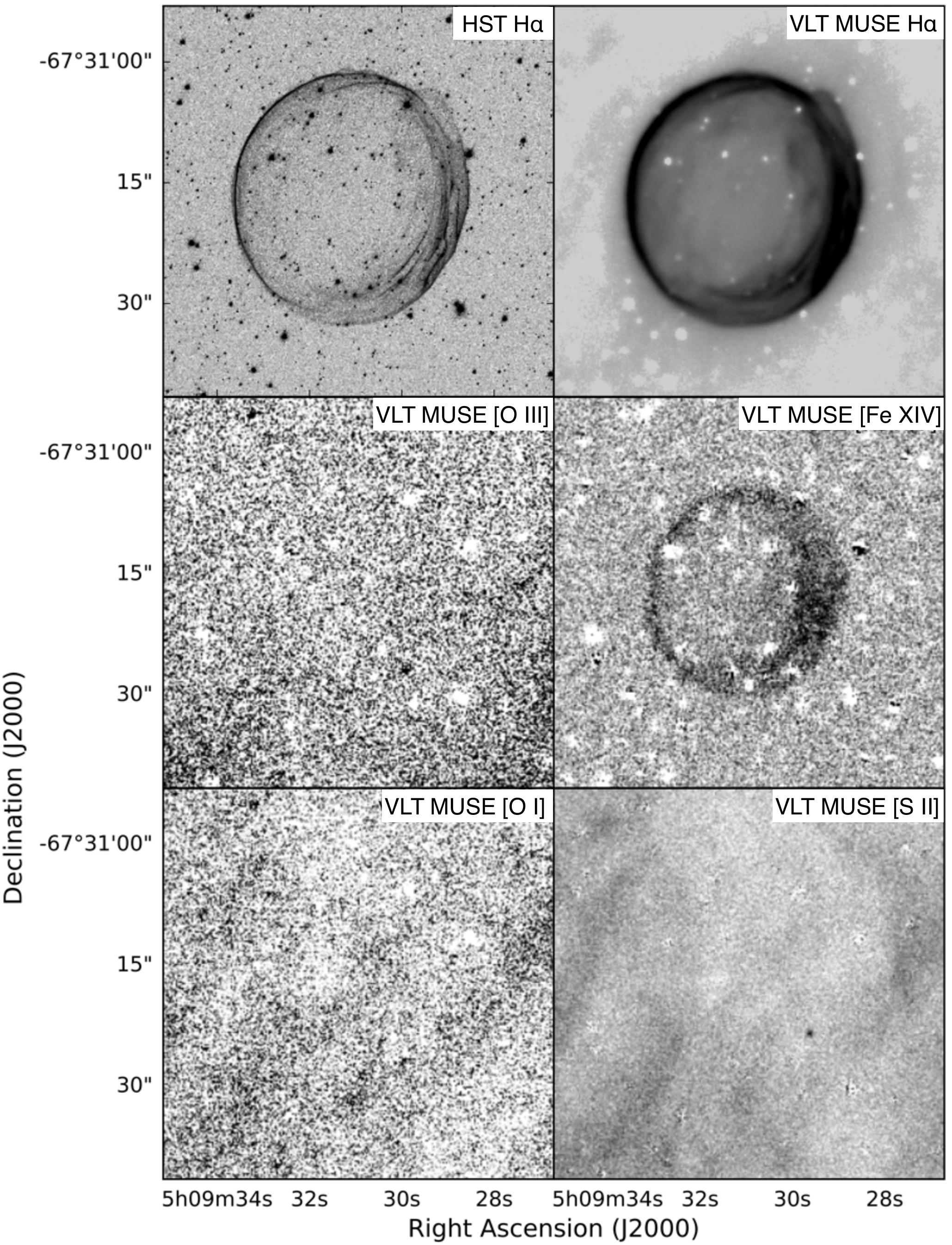}
\end{minipage}%
}%
\hspace{1.5cm}
\subfigure[]{
\hspace{-1.8cm}
\begin{minipage}[t]{0.5\linewidth}
\centering
\includegraphics[scale=0.5]{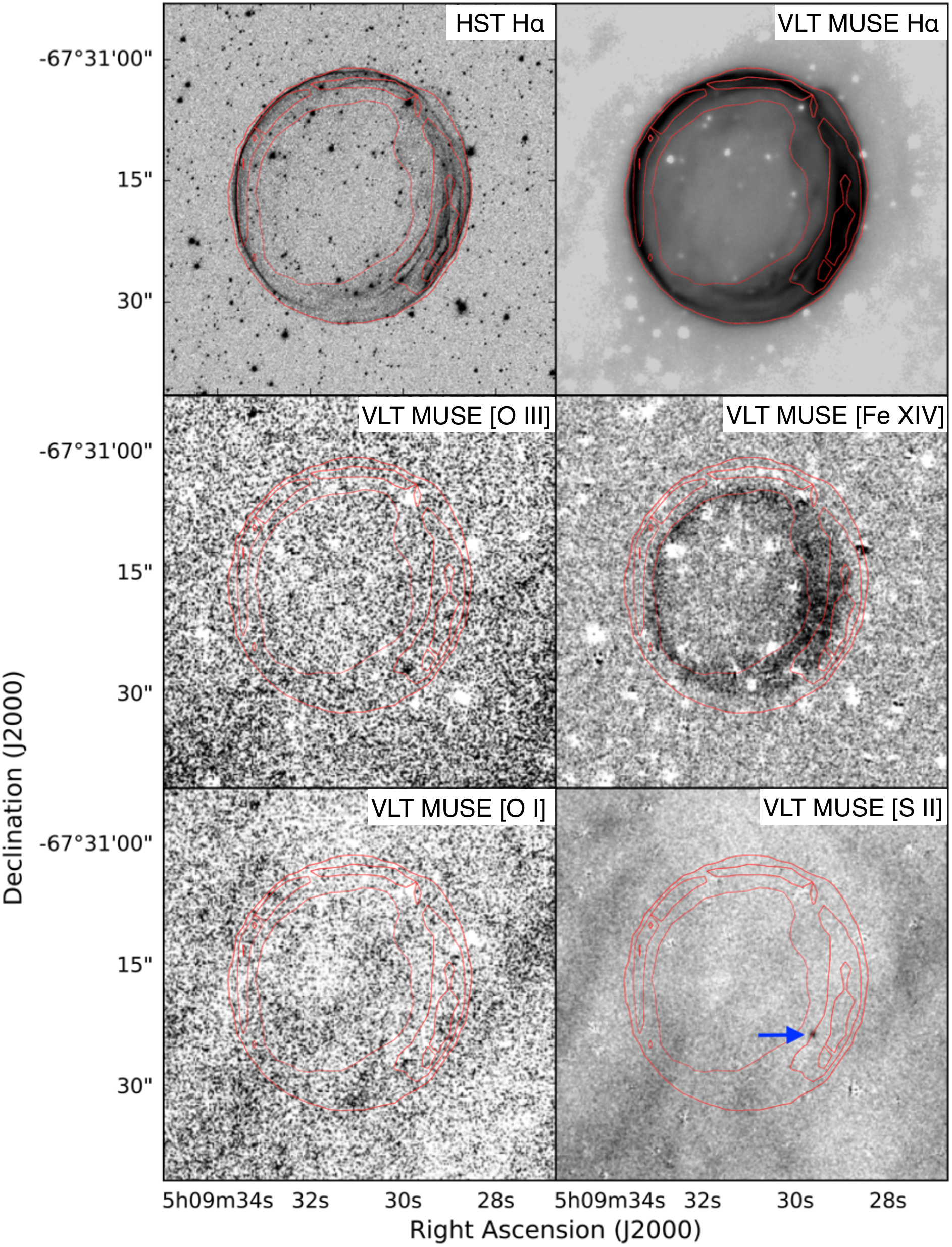}
\end{minipage}%
}%
\caption{(a): \hst \ha and VLT MUSE H$\alpha$, [{\ion{O}{3}}], [{\ion{Fe}{14}}], [{\ion{O}{1}}], and [{\ion{S}{2}}] images of SNR \snr{0509}. (b): Same as panel (a), but with VLT MUSE H$\alpha$\ contours overplotted. In the [{\ion{S}{2}}] image, an blue arrow points to a galaxy that is a residue from the spectral background subtraction.
}
\label{figure:line_imgs_0509}
\end{figure*}

\subsection{SNR \snr{0509} (Figure
\ref{figure:line_imgs_0509})} \label{sec:0509}  


The \hst \ha image of SNR \snr{0509} 
exhibits an overall regular and slightly elliptical shell, as shown 
in Figure \ref{figure:line_imgs_0509}a. The 
eastern side of the shell shows one major filament 
delineating the rim, indicating a simple shell structure, 
while the western side exhibits multiple filaments
along the rim that has been suggested to be caused by a 
non-uniform ambient medium \citep{Hovey2015}. 
The bright filaments originate from locations where the sight line 
is tangent to the shock front, while the diffuse emission near or 
between the filaments originates from locations where the shock front 
is oblique to the line of sight. 
No dense \ha knots are seen in SNR \snr{0509}.

The MUSE \ha image of SNR \snr{0509} detects not only the shell 
rim but also diffuse emission throughout the face of the SNR, 
including the central region where the SNR shocks propagate along 
the lines of sight and the emitting path lengths are the shortest. 
This image also detects a faint \ha halo just outside the SNR shell.  
This halo, not detected in the \hst \ha image, most likely 
originates from the ambient medium
photoionized by the UV precursor of the SNR shocks.
Beyond the halo, to the east and southeast of the SNR exist three 
broad streamer-like features that are most likely from the 
interstellar background and are better seen in the [\ion{S}{2}] image.

The MUSE \oiii line image shows some diffuse emission from the ISM 
background, but no [\ion{O}{3}] emission from the Balmer shell or the 
halo around the shell.
Note that the MUSE spectra show an emission line near $\lambda$5019--5022 \AA,
but this line corresponds to the Doppler shifted \ion{He}{1} $\lambda$5015 line.

The MUSE [\ion{Fe}{14}] line image shows the reverse shock front
into the SN ejecta.  As illustrated in the color composite of 
H$\alpha$, X-ray, and [\ion{Fe}{14}] images in Figure 1 of 
\citet{Seitenzahl2019}, the H$\alpha$ emission from
the forward collisionless shock is located at the largest
radius, the [\ion{Fe}{14}] emission behind the reverse shock
is located at the smallest radius, while the X-ray emission 
is sandwiched between these two shock fronts.

The MUSE \sii line image shows a completely different 
picture from the H$\alpha$ line image.
Neither the filamentary Balmer shell nor the faint diffuse H$\alpha$ 
halo is detected in [\ion{S}{2}].  The most prominent [\ion{S}{2}] 
emission appears in diffuse arcs and streamer-like features on
scales larger than the SNR.
The streamer-like features to the east of the SNR have counterparts 
in the MUSE \ha and [\ion{O}{1}] line images. 
The irregular distribution of these [\ion{S}{2}] features suggests
that they originate from a background ISM.

The MUSE [\ion{O}{1}] line image shows essentially
the same kind of diffuse arcs and streamer-like features that are seen
in the [\ion{S}{2}] image. No [\ion{O}{1}] emission can be unambiguously 
associated with the SNR per se.

The [\ion{S}{2}] line strength relative to the 
respective nearest Balmer lines for the SNR and the interstellar 
background are given in Table~\ref{table:ratio}.
To determine the forbidden line strengths at the Balmer shell rim, 
it is necessary to subtract the background.  
In the [\ion{S}{2}] lines, the background is not only bright 
but also nonuniform.  The [\ion{S}{2}]/H$\alpha$ ratio of the 
filamentary Balmer shell reported in Table~\ref{table:ratio} 
is derived by using a medium to low background value.  If a 
higher background value is used, the [\ion{S}{2}]/H$\alpha$ 
becomes negative. The ``:" symbol denotes a non-detection, 
and the error is dominated by the uncertainties in the background subtraction.
The faint Balmer halo is not detected in [\ion{O}{3}] or 
[\ion{S}{2}], and the large variations in the background ISM  
emission make it impossible to make meaningful estimates of 
their upper limits.   
The [\ion{S}{2}]/H$\alpha$ ratio of the background ISM exhibits 
a range of values, as given in Table~\ref{table:ratio}.
The [\ion{S}{2}] $\lambda$6716/$\lambda$6731 line ratio of the 
background is in the low-density limit, 
$<$100 H-atom cm$^{-3}$, consistent with an ISM origin. 

Note that the emission ``knot" in the southwest quadrant of the SNR in the MUSE \sii
image, marked by an arrow in Figure~\ref{figure:line_imgs_0509}b,
originates from two emission lines at $\lambda$6729 and $\lambda$6734. These two 
lines cannot be the [\ion{S}{2}] doublet from \snr{0509} in the LMC.  Instead, these 
emission lines must be the [\ion{O}{2}] $\lambda\lambda$3726, 3729 lines from a background 
galaxy at z = 0.806, similar to the background galaxy near the center of \snr{0509}
\citep{Litke2017}.  The ratio of the red-shifted [\ion{O}{2}] doublet indicates a density 
near 100 H cm$^{-3}$, consistent with densities of an ISM.  

\begin{figure*}[htbp]
\centering
\hspace{1cm}
\subfigure[]{
\hspace{-2cm}
\begin{minipage}[t]{0.53\linewidth}
\centering
\includegraphics[scale=0.475]{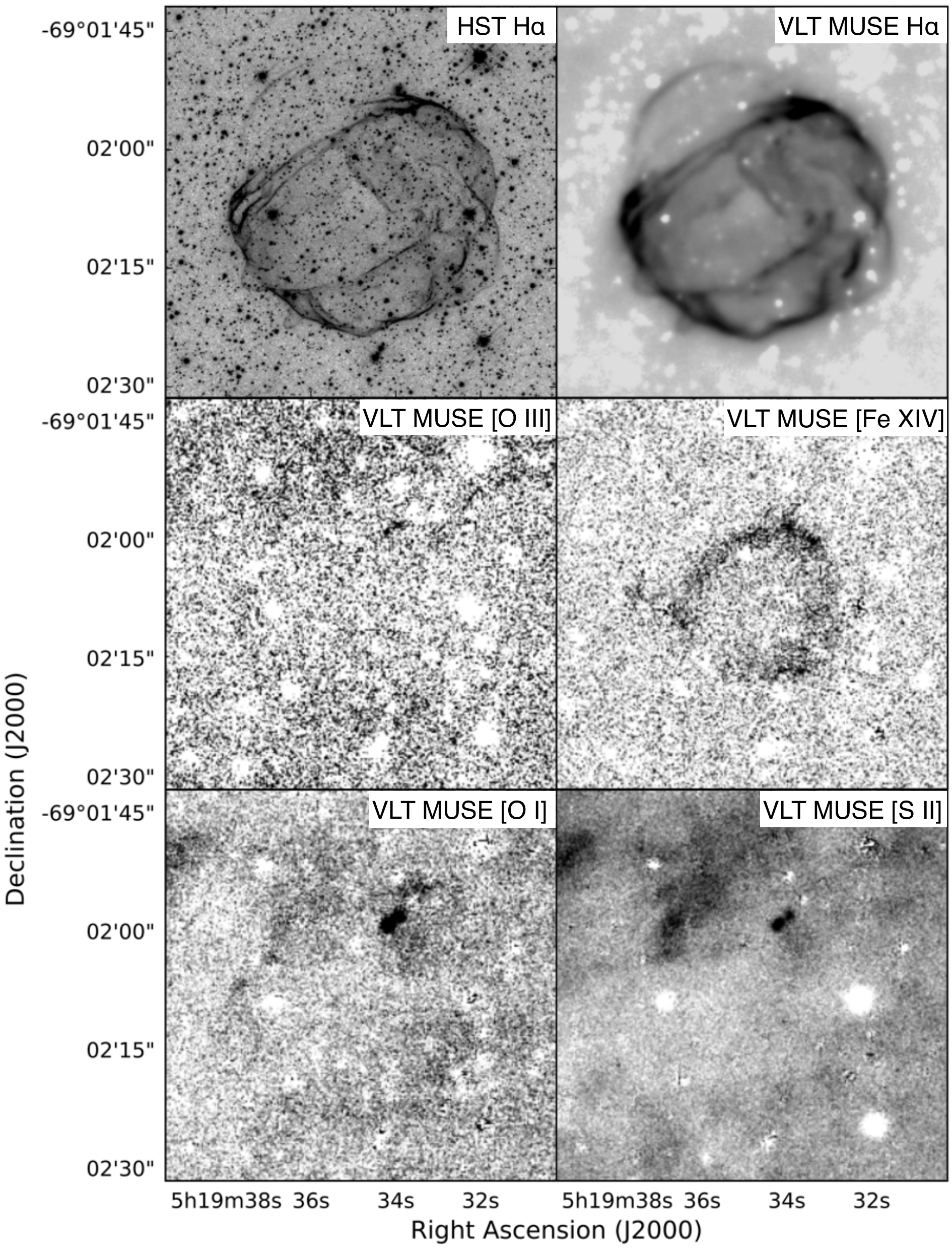}
\end{minipage}%
}%
\hspace{1.5cm}
\subfigure[]{
\hspace{-2cm}
\begin{minipage}[t]{0.5\linewidth}
\centering
\includegraphics[scale=0.5]{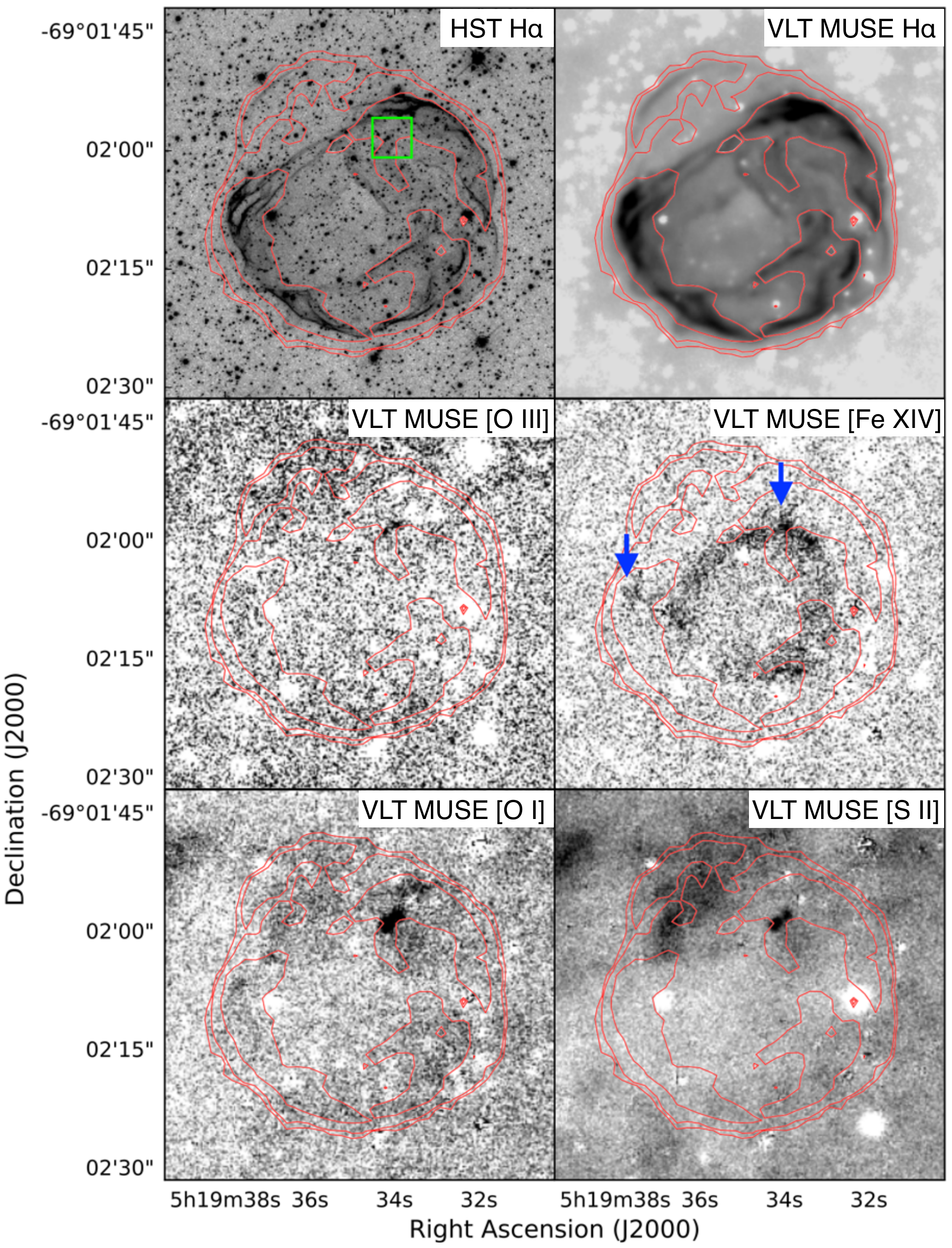}
\end{minipage}%
}%
\caption{Same as Figure \ref{figure:line_imgs_0509}, but for SNR \snr{0519}. The green square marks the region shown in Figure \ref{figure:knots_morphologies}, which illustrates the morphology of knots. See Section \ref{subsubsection:physical}. The blue arrows mark two additional patches of [\ion{Fe}{14}] emission.
}
\label{figure:line_imgs_0519}
\end{figure*}

\subsection{SNR \snr{0519} (Figure
\ref{figure:line_imgs_0519})} \label{sec:0519}  


The \hst \ha image of SNR \snr{0519} shows a more complex 
morphology than that of \snr{0509}.
The SNR shell rim from east through south and west to the
north is roughly round, with morphological features suggestive 
of a ``bubbly'' shell surface.  The northeastern quadrant of the
SNR shell rim appears flattened; however, in this direction 
a faint filamentary arc is detected at larger distances than the 
average shell radius in the other quadrants.  It is likely
that the faint outer arc represents the shock front in the 
northeastern quadrant, while the bright flattened filament and
the bright interior semi-straight filament in the southwestern 
quadrant may have formed by the same mechanism.
No dense knots similar to those in N103B are seen in this \ha image of SNR \snr{0519}.

SNR \snr{0519}'s MUSE \ha line image shows a qualitatively similar 
morphology at a lower resolution than its \hst \ha image 
(see Figure \ref{figure:line_imgs_0519}).
The MUSE \ha image detects emission over the face of the SNR and 
slightly beyond the SNR radius, similar to what is seen in \snr{0509},
although the diffuse \ha halo of \snr{0519} is not as extended.  
This diffuse \ha halo is marginally detected in the \hst \ha image. 
Similarly, this halo most likely originates from an ambient medium 
photoionized by the SNR shock's UV precursor, as in the case of \snr{0509}. 
Beyond the halo, SNR \snr{0519} is surrounded by diffuse,
patchy \ha emission associated with ambient or 
background ISM.

The MUSE \oiii line image reveal a small patch of emission in the 
SNR interior, and this patch is much more prominent in the \sii and 
[\ion{O}{1}] lines images, as shown later in this section.
This \oiii image also shows a very weak partial counterpart of the
Balmer shell: detected from the east through the south to the southwest
along the bright rim and possibly along the faint outer rim in the 
southwest quadrant.  We have carefully separated the \oiii line emission
from the \ion{He}{1} $\lambda$5015 line emission; however, the faint 
high-velocity wings may overlap and contribute to low-level contamination
in limited regions where both lines are detected.
No \oiii counterpart of the faint \ha halo is detected. 
\oiii emission from the diffuse background ISM is detected to the north and
to the south of the SNR.

The [\ion{Fe}{14}] line image shows a shell associated  with the 
reverse shock front into the SN ejecta, similar to that seen in
SNR 0609$-$67.5.  However, 0619$-$69.0 shows two additional patches
of [\ion{Fe}{14}] emission, marked by arrows in Figure \ref{figure:line_imgs_0519}b.
The east patch is projected near the brightest part of the
Balmer shell's eastern rim, while the north patch is projected near 
the one o'clock position of the [\ion{Fe}{14}] ring.  These two small
patches have a different origin from the [\ion{Fe}{14}] shell associated
with the reverse shock, as the FWHM of the line profile is $\sim$300 km s$^{-1}$
in the patches and a few $\times10^3$ kms$^{-1}$ in the shell.
The north patch emits brightly in [\ion{O}{1}] and [\ion{S}{2}] and 
faintly in [\ion{O}{3}]; the east patch shows [\ion{O}{3}] and 
[\ion{O}{1}] emission, but it is difficult to disentangle the 
contributions from the Balmer shell and the east patch.

The \sii line image reveals two distinct knots of bright emission 
projected within the SNR shell, about half way from the shell center 
to the bright H$\alpha$ shell rim in the north.
These \sii knots are coincident with the north patch of [\ion{Fe}{14}]
emission, and have faint counterparts in 
the \oiii line image, too.  In the MUSE \ha image, these 
knots blend in with Balmer filaments and do not appear as distinct 
features; however, the \hst \ha image reveals that
the two knots are connected by diffuse emission of lower
surface brightness and the knots have diameters of 
$\lesssim$ 0\farcs3 (0.075 pc).
These nebular knots are so small that they easily
escape detection if the \hst \ha image alone is
examined.  The presence of these isolated small knots 
is very intriguing and their origin will be discussed
later in this paper.  Besides these knots, no [\ion{S}{2}]
emission can be unambiguously associated with the Balmer shell
or the faint \ha halo,
although an irregular diffuse ISM emission background is
clearly detected.

The [\ion{O}{1}] line image shows not only bright counterpart 
of the north [\ion{Fe}{14}] patch that is coincident with 
the [\ion{S}{2}] knots, but also enhanced emission at the east 
[\ion{Fe}{14}] patch that is not detected in [\ion{S}{2}].
In addition, [\ion{O}{1}] emission is seen along the southern half of
the Balmer shell rim.  There is enhanced [\ion{O}{1}] emission 
enclosed within the northern part of the Balmer shell, but it
is not clear whether the [\ion{O}{1}] emission is associated with the
SNR shell or the background diffuse ISM.

The [\ion{O}{3}] and [\ion{S}{2}] line emission from the Balmer shell and the 
background ISM of \snr{0519} is measured in a similar 
fashion as that of \snr{0509}.
The two dense knots in \snr{0519} are measured with very 
high signal-to-noise ratios.
The [\ion{S}{2}]/H$\alpha$ and [\ion{O}{3}]/H$\beta$ ratios 
of the Balmer shell, dense knots, and the background ISM are 
given in Table~\ref{table:ratio}.
The [\ion{S}{2}] doublet ratio of the background is in the 
low density limit and consistent with an ISM origin.



\begin{figure*}[htbp]
\centering
\hspace{1cm}
\subfigure[]{
\hspace{-2cm}
\begin{minipage}[t]{0.53\linewidth}
\centering
\includegraphics[scale=0.5]{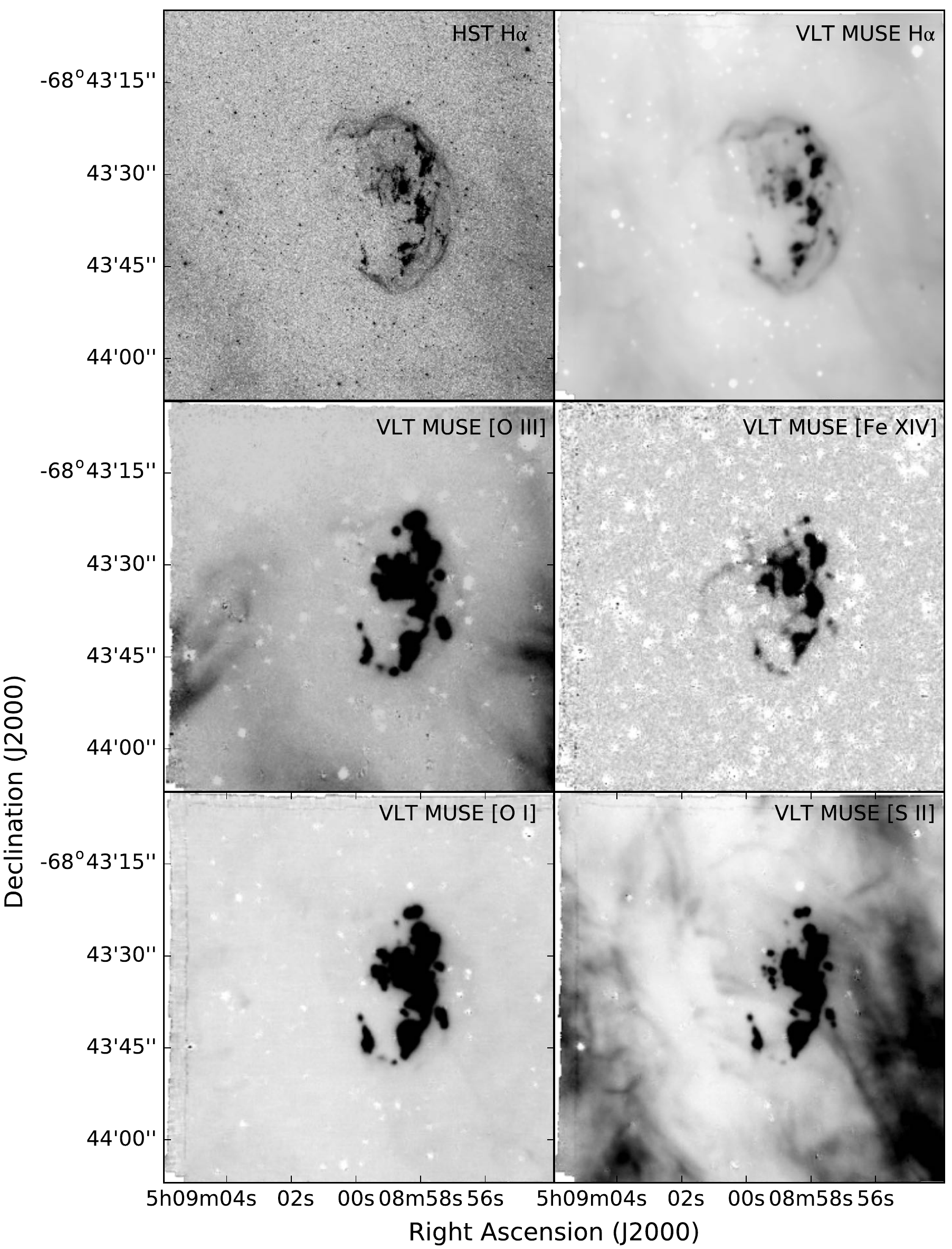}
\end{minipage}%
}%
\hspace{1.5cm}
\subfigure[]{
\hspace{-2cm}
\begin{minipage}[t]{0.5\linewidth}
\centering
\includegraphics[scale=0.476]{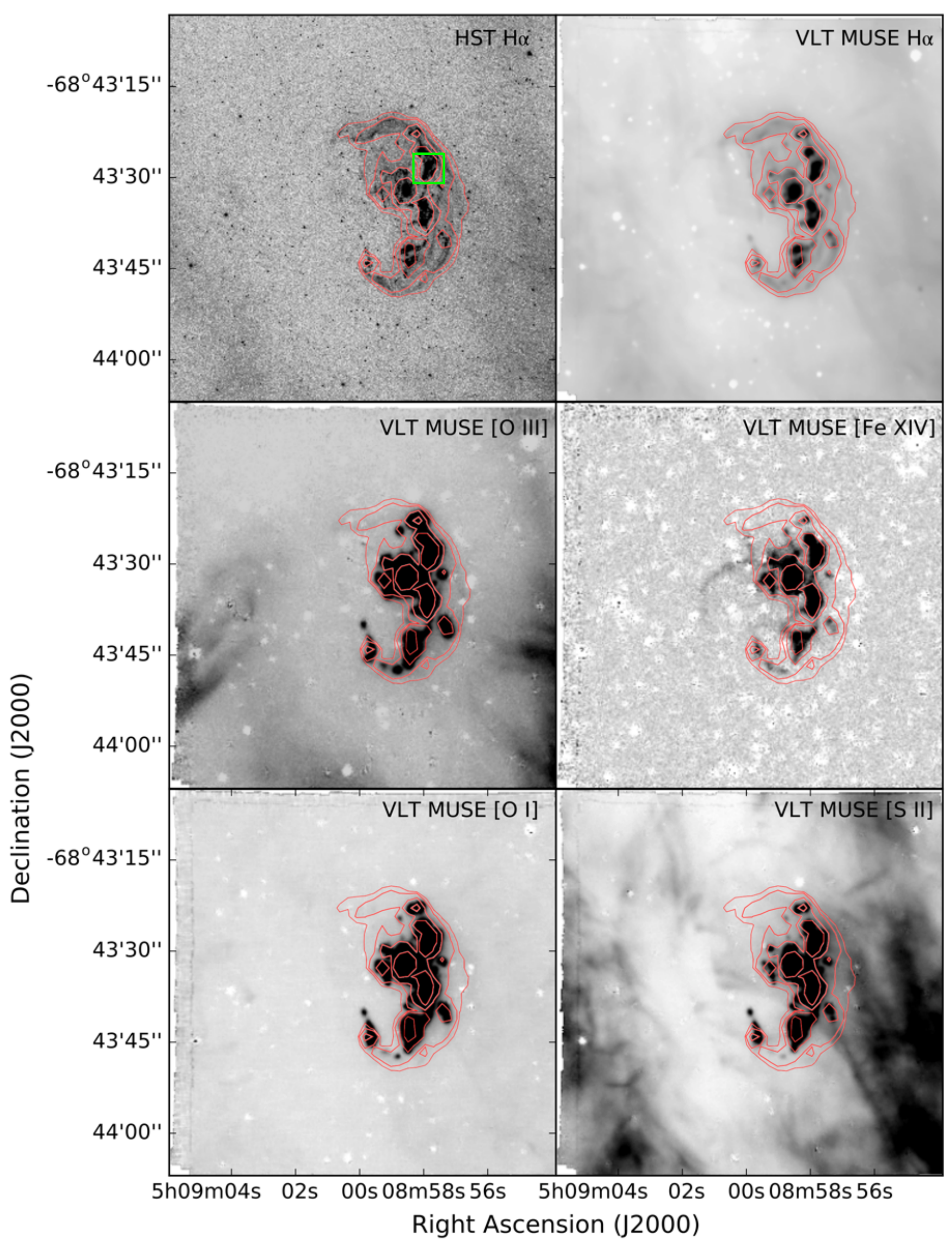}
\end{minipage}%
}%
\caption{Same as Figure \ref{figure:line_imgs_0519}, but for SNR N103B.}
\label{figure:line_imgs_n103b}
\end{figure*}

\subsection{SNR N103B (Figure
\ref{figure:line_imgs_n103b})} \label{sec:n103b}  

The \hst \ha image of SNR \snr{n103b} clearly reveals a 
filamentary shell structure that encompasses prominent 
groups of dense knots.  Only the filamentary shell is 
Balmer-dominated, and the dense knots represent a 
CSM ejected by the SN progenitor \citep{Williams2014,Li2017}.
The shell is elliptical with an opening to the east. 
The knots are mainly distributed in four groups, with 
only few knots located along \ha filaments near the 
shell rim. Almost all knots are seen on the western 
side of the shell interior.
The knots individually can have sizes as small as  
0\farcs14 (0.035 pc) in diameter, while a group of knots
can be as extended as 2$''$ (0.5 pc) across.

N103B's MUSE \ha line image shows bright nebular 
knots in the interior of an incomplete shell, 
similar to those seen in the \hst \ha image, except 
the ``knots'' seen in the MUSE image correspond
to ``groups of knots'' in the \hst image.
The MUSE \ha image also shows that N103B is 
projected in a complex diffuse background with
multiple filaments and arcs in the surroundings.

The MUSE \oiii line image shows bright emission 
from the knots, and very faint emission from the 
Balmer-dominated shell. The surrounding ionized
ISM is also detected in the \oiii image with a
surface brightness distribution following that 
of the \ha image roughly but not exactly.
The [\ion{O}{3}]/\ha ratio varies among the CSM knots and
in the surrounding ISM; their values are given in
Table \ref{table:ratio}.

The MUSE [\ion{Fe}{14}] line image shows bright emission from
the CSM knots and an arc extending to the east.  This arc
has no counterpart in other optical line images; however, it
follows an X-ray arc closely \citep{Seitenzahl2019}.  
The FWHM of the [\ion{Fe}{14}] line profile is up to 3300 km s$^{-1}$
in the arc, and 400 km s$^{-1}$ in the knots.  These widths and
spatial distributions are consistent with the origin of reverse 
shock into the SN ejecta for the arc, and forward shock into the 
CSM for the knots.

The MUSE \sii line image shows bright emission from
the knots, but no emission from the Balmer-dominated 
shell.  
The \sii emission from the surrounding ISM follows
\ha emission well. The \sii image also shows two
intriguing concentric ring-like structures: the inner
ring has a radius of 4.9 pc and roughly encompasses 
the SNR, while the outer ring is 5.9 pc in size
and is projected in a background with low
\ha surface brightness. 

The MUSE [\ion{O}{1}] image shows bright emission from the
CSM knots.  The background emission from diffuse ISM is also
detected at a much lower level than the CSM knots.

In summary, N103B has prominent forbidden line 
emission from the CSM knots, faint \oiii line
emission from the Balmer-dominated shell, and 
moderately bright forbidden line emission from
the surrounding ISM.  The representative ranges of
forbidden line strengths from different features
in and around the SNR N103B are given in 
Table~\ref{table:ratio}.
The dense knots' electron densities determined 
from the [\ion{S}{2}] doublet are 10$^3$--10$^4$
cm$^{-3}$, consistent with the $3\times10^4$--$10^5$ cm$^{-3}$
determined from the UV lines of \ion{Si}{3} \citep{Blair2020},
and their electron temperatures determined 
from the [\ion{N}{2}] $\lambda$6583 and 
$\lambda$5577 lines are 10,000 -- 20,000 K.
The electron density of the background is below 
100 cm$^{-3}$, consistent with an ISM origin.

\begin{figure*}[htbp]
\centering
\hspace{1.cm}
\subfigure[]{
\hspace{-2cm}
\begin{minipage}[t]{0.53\linewidth}
\centering
\includegraphics[scale=0.5]{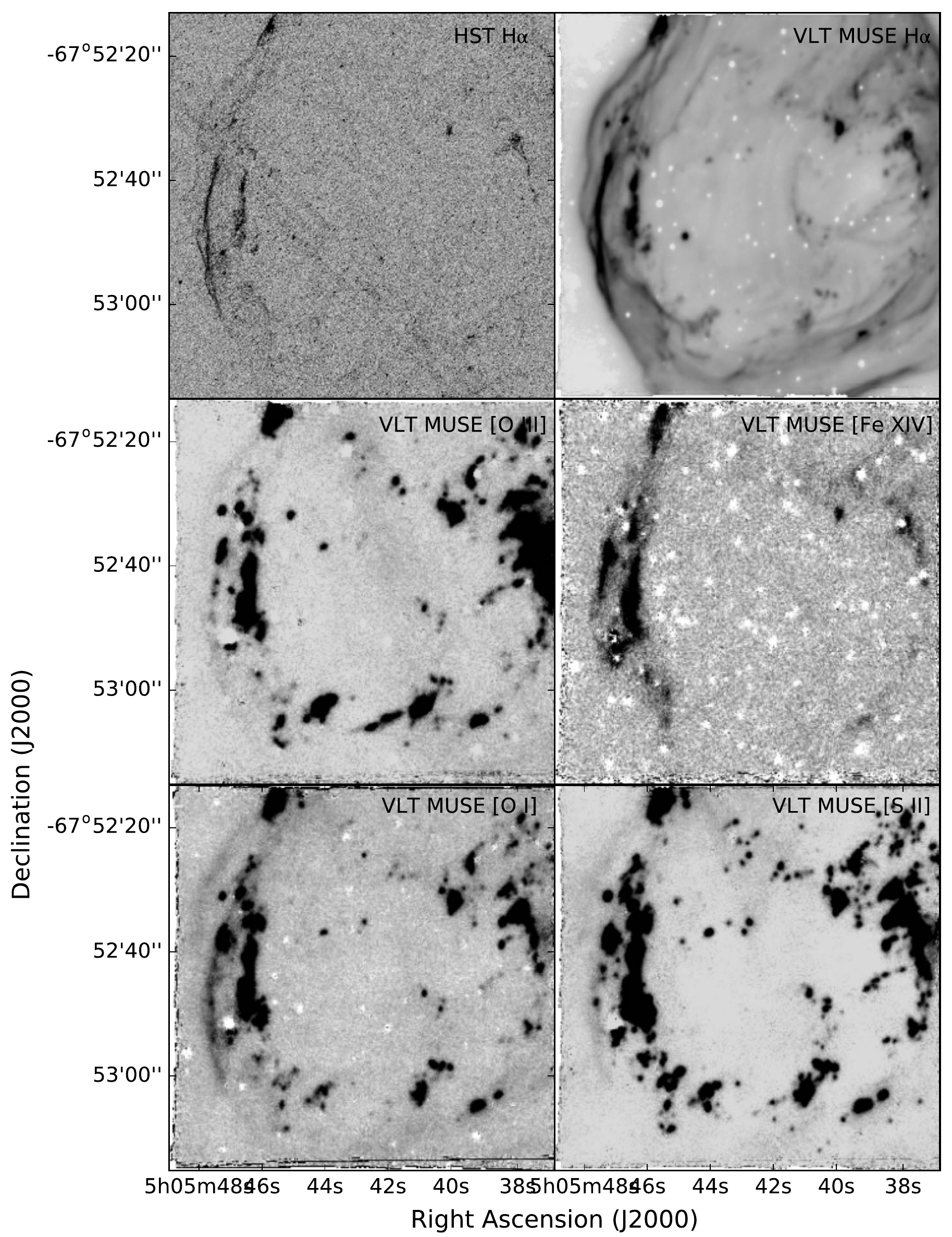}
\end{minipage}%
}%
\hspace{1.5cm}
\subfigure[]{
\hspace{-2cm}
\begin{minipage}[t]{0.5\linewidth}
\centering
\includegraphics[scale=0.476]{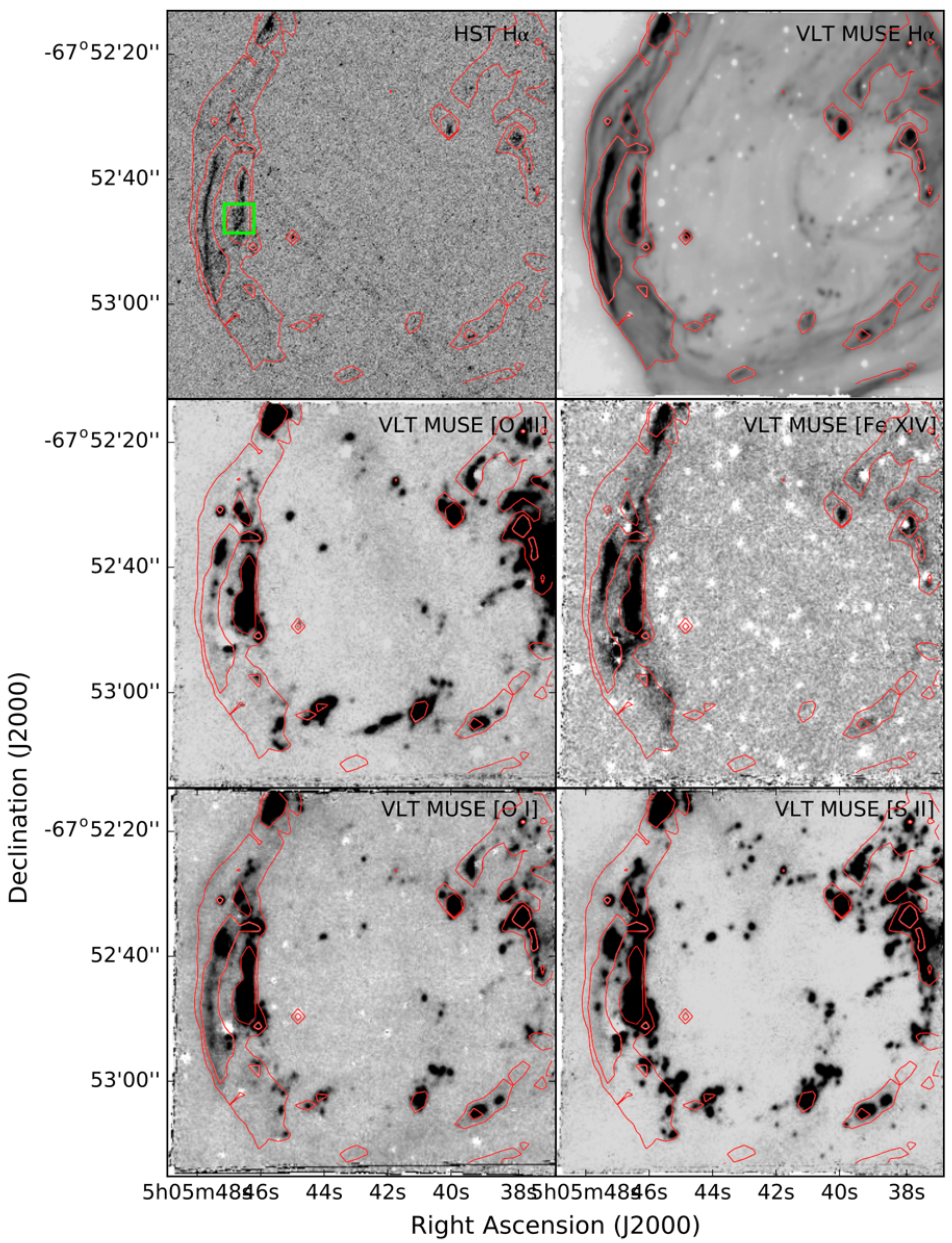}
\end{minipage}%
}%
\caption{Same as Figure \ref{figure:line_imgs_0519}, but for SNR DEM L71.}
\label{figure:line_imgs_deml71}
\end{figure*}

\subsection{SNR DEM L71(Figure
\ref{figure:line_imgs_deml71})} \label{sec:deml71}  
 

The \hst \ha image shows that \snr{deml71}'s SNR shell is quite irregular. 
The eastern and southwestern sides of the shell contain
interior filaments curved along the outer rim.
The northern and southwestern parts of the shell rim
bulge out, suggestive of blow-outs into a lower-density
medium.  Upon closer inspection, small knots can be seen
distributed along some interior filaments and outer rim.  The smallest nebular knots have diameters $\lesssim$ 0\farcs3 (0.075 pc).  A small number of nebular knots 
are projected in the shell interior, 
and they appear diffuse with lower surface 
brightness.

\snr{deml71}'s MUSE \ha line image has a 
smaller field of view than the \hst
\ha image. In the MUSE \ha image, the eastern
part of the main SNR shell is clearly detected
but the shell rims in the other directions are 
outside the field of view. The MUSE \ha 
image, being much deeper than the \hst \ha image, 
reveals numerous filaments within the SNR
shell, especially those curved along the shell 
rim from the east through the south to the west. 
In addition, some \ha nebular knots are detected.

The MUSE \oiii line image shows a very different
morphology: bright knots are detected along the
aforementioned curved filaments interior to the 
shell, and some knots are projected in the central
cavity without association with filaments.  
The \oiii image also detects faint counterpart
of the eastern rim of the Balmer-dominated shell.
Also detected is a band of diffuse emission extending
from the top center southwestwards to the bottom 
of the field of view. 

The MUSE [\ion{Fe}{14}] line image shows emission 
from the knots near the east rim and northwest rim.
In addition, [\ion{Fe}{14}] emission is detected 
along the Balmer shell rim on the east
and west sides of the SNR.  It is interesting that 
the [\ion{Fe}{14}] arcs follow X-ray emission that
delineates the forward shocks into the ISM.  This is 
quite different from what we see in 0509$-$67.5, 
0519$-$9.0, and N103B.

The MUSE \sii line image shows nebular knots even more 
prominently than the \oiii image, detecting more faint 
knots.  The MUSE \sii image also detects faint emission 
from the Balmer-dominated shell rim on the east.  
The \sii image of \snr{deml71} has also detected 
the northern part of the band of diffuse emission 
seen in the \oiii image.

The MUSE [\ion{O}{1}] line image shows emission from
both knots and the Balmer shell rim on the east side.
The [\ion{O}{1}] image is qualitatively similar to the
[\ion{S}{2}] image, although the [\ion{O}{1}] emission 
along the Balmer shell rim is more prominent.


\begin{figure*}[htbp]
\centering
\hspace{0.8cm}
\subfigure[]{
\hspace{-2cm}
\begin{minipage}[t]{0.55\linewidth}
\centering
\includegraphics[scale=0.5]{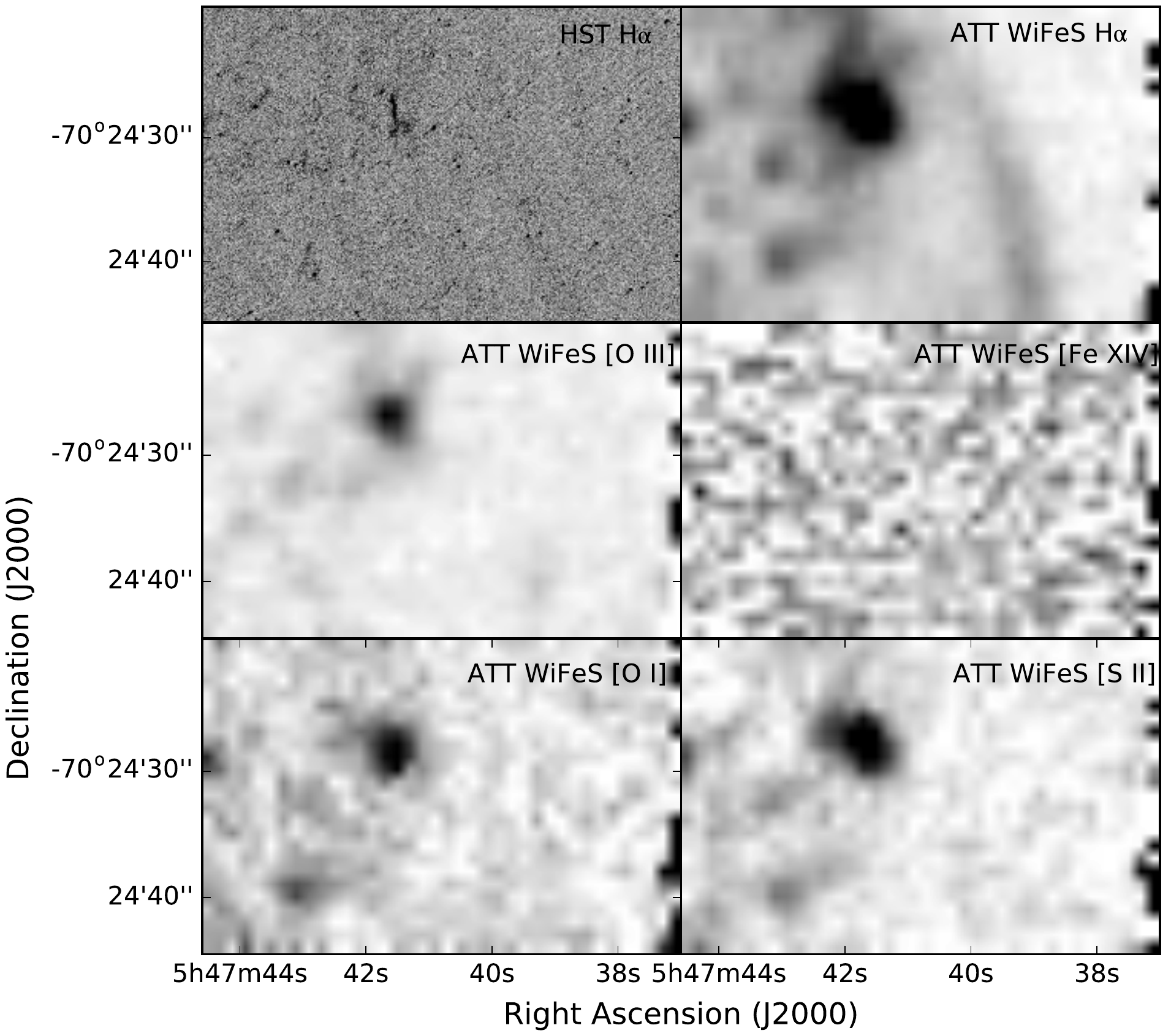}
\end{minipage}%
}%
\hspace{1.5cm}
\subfigure[]{
\hspace{-2cm}
\begin{minipage}[t]{0.5\linewidth}
\centering
\includegraphics[scale=0.465]{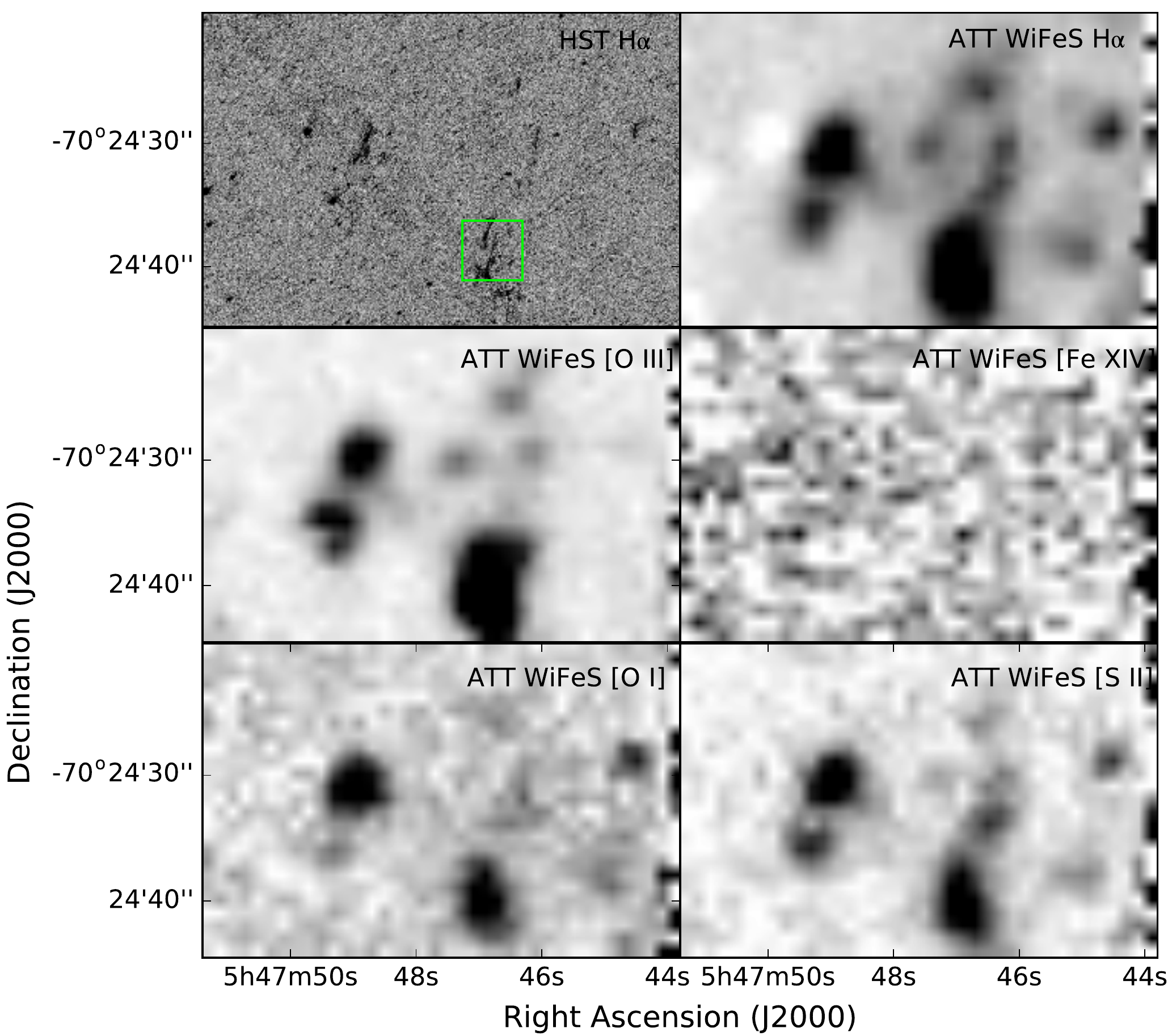}
\end{minipage}%
}%
\caption{
(a): \hst \ha and ATT WiFeS H$\alpha$, [{\ion{O}{3}}], [{\ion{Fe}{14}}], [{\ion{O}{1}}], and [{\ion{S}{2}}] images for the G9 data of SNR \snr{0548}. (b): 
Same as panel (a), but for the G4 data of SNR \snr{0548}.  The green square marks the region shown in Figure \ref{figure:knots_morphologies}, which illustrates the morphology of knots.
}
\label{figure:line_imgs_0548}
\end{figure*}





\begin{figure*}
\epsscale{1.2}
\plotone{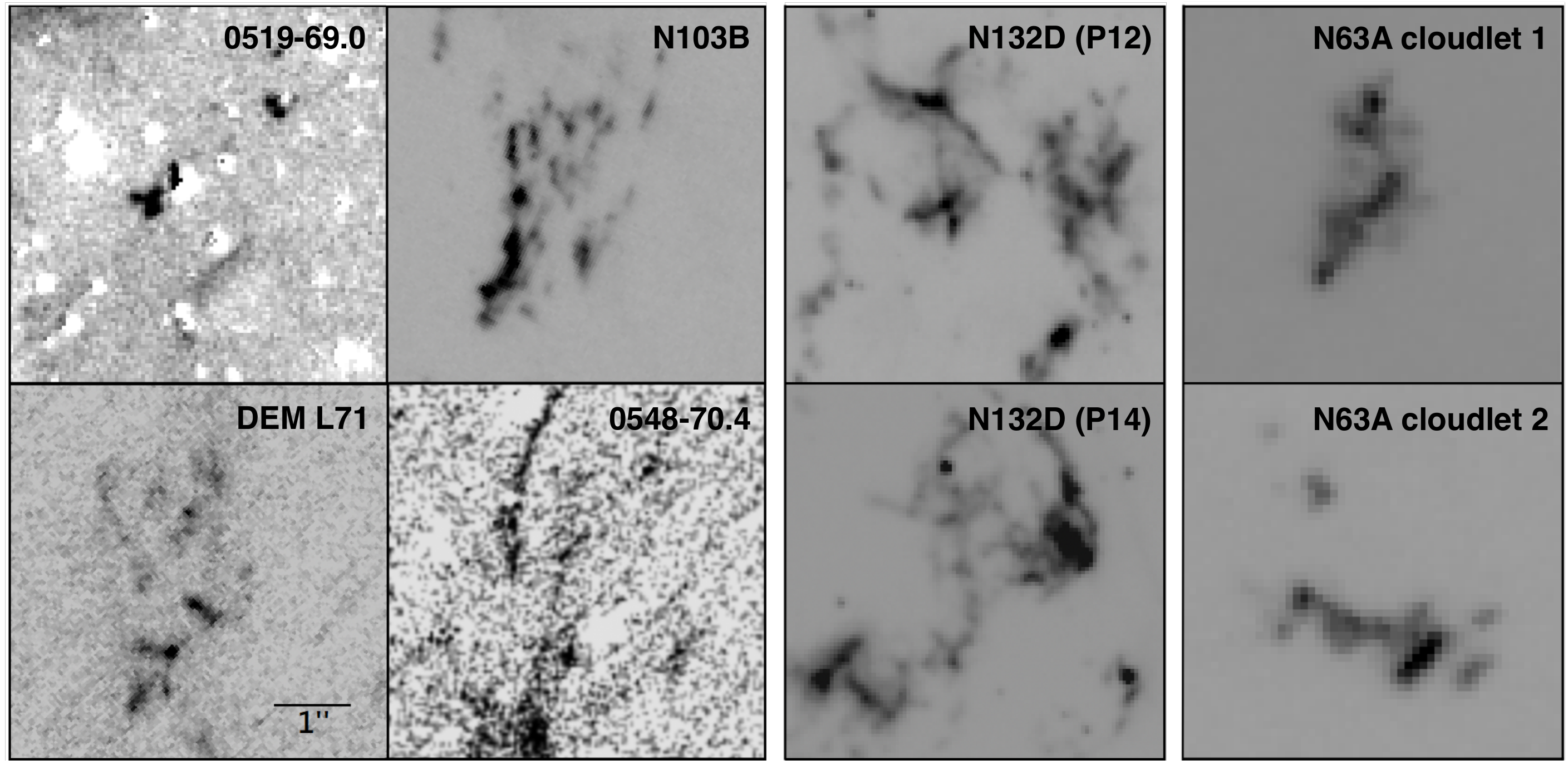}
\caption{ 
The \hst \ha images showing different morphologies of knots in four SNRs \snr{0519}, N103B, DEM\,L71, \snr{0548}. The continuum emission has been subtracted from the \snr{0519} and \snr{0548} \ha images. The field of view of each panel is 5\farcs0 $\times$ 5\farcs0. These regions are marked with green squares in Figures \ref{figure:line_imgs_0519}b, \ref{figure:line_imgs_n103b}b, \ref{figure:line_imgs_deml71}b, and \ref{figure:line_imgs_0548}b. The P12 and P14 regions of SNR N132D in \citet{Dopita2018} and cloudlets of N63A in \citet{Chu1999} are also shown for comparisons. 
}
\label{figure:knots_morphologies}
\end{figure*}

The MUSE observation of \snr{deml71} covered only
a small area of sky outside the SNR shell. 
While this sky coverage is adequate for background 
subtraction from bright features of small sizes, such
as the Balmer shell rim and knots, it is inadequate 
for accurate measurements of forbidden line emission
from the background per se or background subtraction
for extended features, such as the band of diffuse 
emission across the face of \snr{deml71}.
Therefore, only the Balmer shell and the knots are
reported in Table~\ref{table:ratio}.
The electron densities of the knots are in the range
from 650 to over 10,000 cm$^{-3}$, and the electron temperatures
are $\sim$13,000 K. 
The electron density derived from the [\ion{S}{2}] doublet
of the SNR shell rim is in the low-density limit,
$<$100 cm$^{-3}$.


\subsection{SNR \snr{0548}(Figures
\ref{figure:line_imgs_0548}a and \ref{figure:line_imgs_0548}b)} \label{sec:0548}  

The \hst \ha image of SNR \snr{0548} has the lowest 
signal-to-noise ratio compared to similar observations
of \snr{n103b} and \snr{deml71}, because of the relatively
low surface brightness of \snr{0548}.  The \bd shell of 
\snr{0548} appears quite regular (see Figure \ref{figure:Ha_imgs_5SNRs_new_2}),
although it shows double-rim morphology along the
northeastern quadrant, similar to those seen in \snr{deml71}.
Projected near the central region of \snr{0548} are four
strips of diffuse emission with some of them containing 
nebular ``rods'' that are 0\farcs25 (0.063 pc) in width and
up to 2$''$ (0.5 pc) in length.  There is also enhanced
nebular emission, possibly similar to knots or rods, along 
a filament near the northwestern rim.

The ATT WiFeS observations of SNR \snr{0548} mapped the entire remnant (Fig.\ \ref{figure:SNR0548_pretty}); 
however, only data of the G4 and G9 fields are of adequate 
quality to extract spectral information.  
The G9 field includes a segment of the Balmer-dominated 
shell rim and one of the four aforementioned strips of
diffuse emission in the central region of the SNR
(Fig.~\ref{figure:line_imgs_0548}a).
The [\ion{O}{3}], [\ion{O}{1}], and \sii line images 
detect emission from knots and the strip of diffuse 
emission, but the Balmer-dominated shell rim is not
unambiguously detected in any of the forbidden line images.
The [\ion{Fe}{14}] line is not detected anywhere.
The G4 field includes strips of diffuse emission with 
embedded knots (Fig.~\ref{figure:line_imgs_0548}b).  
These features are detected in [\ion{O}{3}], [\ion{O}{1}], 
and \sii line images.  The [\ion{Fe}{14}] line is not
detected.
The relative strengths of the [\ion{O}{3}], [\ion{O}{1}],
and \sii lines vary among the emission regions. 
The ranges of representative \oiiitohb and 
\siitoha ratios are given in Table \ref{table:ratio}.




 


\begin{deluxetable*}{ccccccccc}
\tablecaption{ Representative physical quantities and properties in and around SNRs with Balmer-dominated shells }
\tablehead{
SNR$^a$ & Age  & Features & \siitoha &  \oiiinospace$/$\hb & n$_{\textrm{e}}$  & T   \\
 & (yr) &  &  &  & (cm$^{-3}$) & (K) 
}
\startdata 
\snr{0509}  &  400$\pm$50$^b$ & Filamentary Shell & $<$ 0.004 $\pm$ 0.032: &  ... & ... &  ...  \\
... & ... & Background ISM & (0.9 -- 1.5) $\pm$ 0.1 & ... &  $<$ 100 & ...  \\
\hline
... & ... & Filamentary Shell & $<$ 0.026 $\pm$ 0.041: &  $\leq$ 0.017 $\pm$ 0.003 &  ...  &  ... \\ 
\snr{0519} & 600$\pm$200$^b$ & Bright Knots &   (0.2 -- 1.0) $\pm$ 0.0 &  (0.2 -- 1.5) $\pm$ 0.2 &  1,500 -- $\ge$10,000 & ...  \\
... & ... & Background ISM &  (0.5 -- 0.9) $\pm$ 0.2 &  (0.0 -- 0.8) $\pm$ 0.1  & $<$ 100  &  ... \\
\hline
... & ... & Filamentary Shell & $<$ 0.062 $\pm$ 0.130: & $\leq$ 0.093 $\pm$ 0.011  & ... &  ... \\
\snr{n103b} & 860$^b$ & Bright Knots &  (0.1 -- 0.6) $\pm$ 0.0 &  (0.2 -- 2.6) $\pm$ 0.0 & 1,000 -- $\ge$10,000 & 9,000 -- 19,000   \\
... & ... & Background ISM &  (0.1 -- 0.5) $\pm$ 0.1 &  (0.1 -- 0.9) $\pm$ 0.1 & $<$ 100 &  ...  \\
\hline
... & ... & Filamentary Shell & $<$ 0.034 $\pm$ 0.016 & $\leq$ 0.043 $\pm$ 0.005   & $<$ 100 & ... \\
\snr{deml71} & 4,360$\pm$290$^c$ & Bright Knots &  (0.2 -- 1.3) $\pm$ 0.0 &  (0.2 -- 5.5) $\pm$ 0.0 & 650 -- $\ge$10,000 & 11,000$^d$  \\
... & ... & Background ISM & ... &  ... & ... & ... \\
\hline
... & ... &  Filamentary Shell &  $<$ 0.03 $\pm$ 0.01: & $\leq$ 1.15 $\pm$ 0.13 & ... & ... \\
 \snr{0548} & $\sim$10,000$^e$ & Bright Knots & (0.3 -- 0.7) $\pm$ 0.0 & (1.9 -- 5.6) $\pm$ 0.4 & 400 -- 2,500 & ... \\
... & ... &  Background ISM & (0.1 -- 0.5) $\pm$ 0.1 &  (2.1 -- 2.6) $\pm$ 0.6 &  $<$ 100 & ...
\enddata
\tablenotetext{a}{  Integrated H$\alpha$ fluxes measured from the MUSE data are: 
$1.5\times10^{-13}$, $2.9\times10^{-13}$, $1.2\times10^{-12}$, and $>6.7\times10^{-13}$ 
erg s$^{-1}$ cm$^{-2}$ for 0509$-$67.5, 0519$-$69.0, N103B, and DEM\,L71, respectively. Note
that the MUSE field is too small to cover the entire DEM\,L71 and the H$\alpha$ flux of DEM\,L71
represents only a lower limit.
} 
\tablenotetext{b}{ From \citet{Rest2005}. } 
\tablenotetext{c}{ From \citet{Ghavamian2003}. }
\tablenotetext{d}{ Very few knots have \nii $\lambda$5755 line clearly detected. }
\tablenotetext{e}{ From \citet{Hendrick2003}. }
\label{table:ratio}
\end{deluxetable*}

\section{Discussion} \label{sec:Discussion}  

We have found three sources of forbidden line emission in
Type Ia SNRs with Balmer-Dominated Shells: bright emission from
dense knots inside the SNRs, [\ion{Fe}{14}] emission from
reverse shocks into the SN ejecta, and faint emission associated
with the collisionless shocks.  The physical properties and
origins of these sources are discussed below.

\subsection{Dense Knots}

\subsubsection{Physical Properties of the Knots}
\label{subsubsection:physical}

The content and distribution of dense knots are different among the five 
Type Ia SNRs with Balmer-Dominated Shells we studied: in ground-based [\ion{S}{2}]
images, \snr{0509} has no knots at all, 
\snr{0519} has two knots close together in the northwest 
quadrant of the SNR, N103B has prominent groups of knots distributed
only in the west side of the SNR, DEM\,L71 has bright knots distributed 
along inner Balmer filaments in a ring-like structure with fainter knots 
projected interior to the ring, and \snr{0548} has knots distributed mostly
within a few patches projected near the central region of the SNR. 

The ``knots" detected in ground-based images are often
resolved into multiple smaller knots in HST images.
The morphologies of these small knots are different among the SNRs.  
HST H$\alpha$ images show that 
some knots are round or slightly elongated with minor axis as small as
$\sim$0\farcs2 and major-to-minor axis ratio ranging from 1.0 to a few, as seen
in N103B and DEM\,L71, while some knots are rod-like with widths of $\sim$0\farcs2
and lengths of 2$''$--3$''$, as seen in \snr{0548}.  The knots in \snr{0519}
are best detected in the MUSE [\ion{O}{1}] and \sii line images where two knots are seen.  
The HST H$\alpha$ image resolved these knots into knots as small
as 0\farcs3 (0.075 pc) in diameter and connecting filaments.  
The different morphologies of knots are illustrated in Figure \ref{figure:knots_morphologies}.

The spectral properties vary among the knots as well.  The variations can be
easily seen from comparisons among the line images.  For example, N103B and DEM\,L71
have knots that are well detected in [\ion{O}{3}] but appear very faint in [\ion{S}{2}],
and vice versa.  Most interestingly, of the two [\ion{Fe}{14}] knots in \snr{0519} 
(marked in Figure \ref{figure:line_imgs_0519}, one 
is bright in [\ion{S}{2}] and [\ion{O}{1}] but faint in [\ion{O}{3}], while the other 
has faint counterparts in [\ion{O}{1}] and [\ion{O}{3}] but not detected in [\ion{S}{2}]. 

Electron densities of the knots, determined from the \sii $\lambda$6716/$\lambda$6731 
ratios, range from a few $\times 10^2$ to $10^4$ cm$^{-3}$.  \snr{0548} has
the lowest density in the knots, which also have the rod-like morphology.

\begin{figure*}
\epsscale{1.3}
\plotone{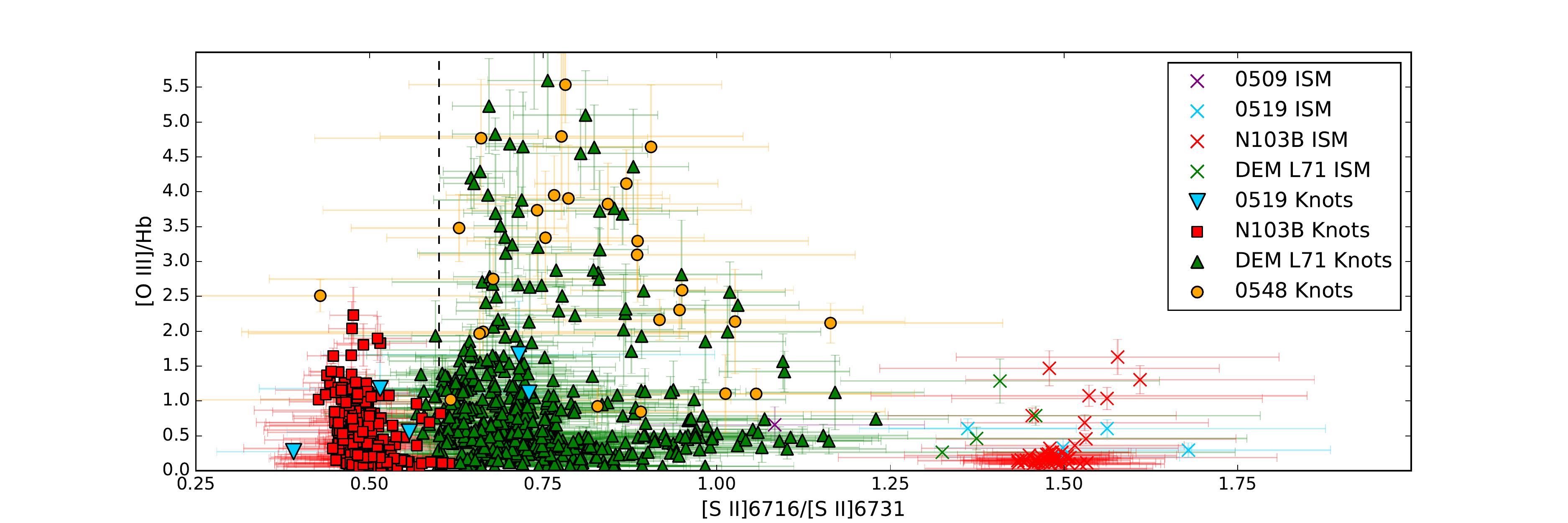}
\caption{ 
The \oiiitohb versus \sii doublet ratio plot in and around LMC Type Ia SNRs with Balmer-dominated shells.
}
\label{figure:oiii_hb_sii_ratio_Dec28}
\end{figure*}

\begin{figure*}
\epsscale{1.3}
\centering
\plotone{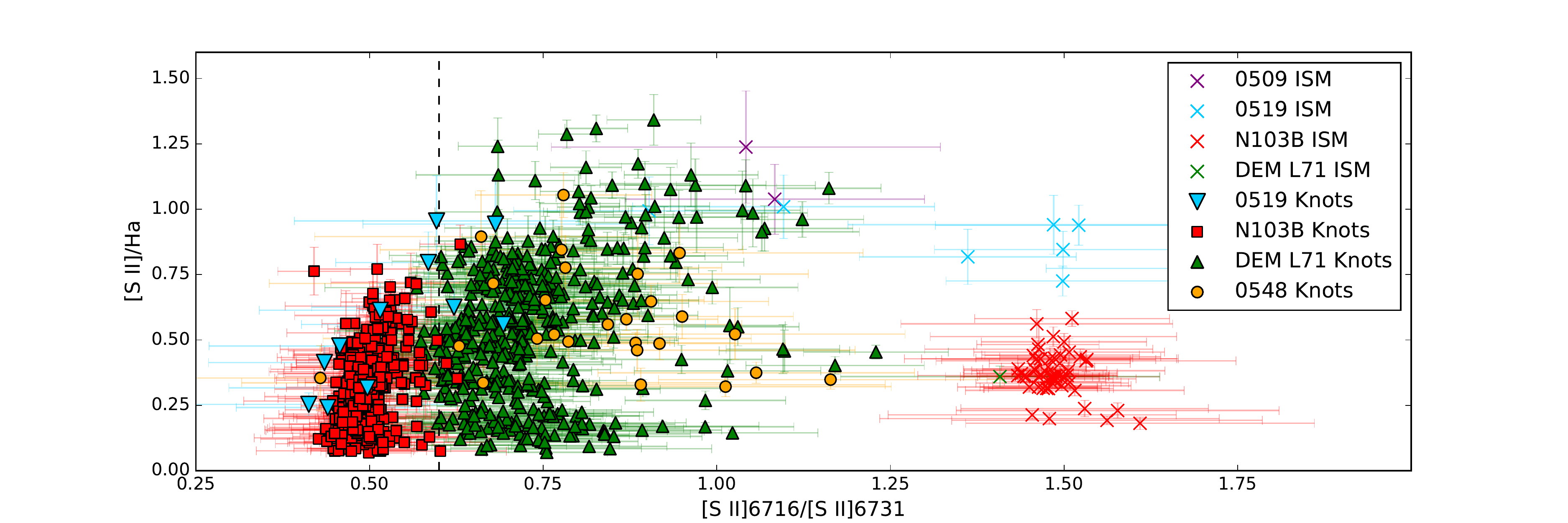}
\caption{
The \siitoha versus \sii doublet ratio plot in and around LMC Type Ia SNRs with Balmer-dominated shells.
}
\label{figure:sii_ha_sii_ratio_Dec28}
\end{figure*}

The [\ion{O}{3}]/H$\beta$ ratios of the knots range from 0.2 to 5.6: 
\snr{0519} and N103B have lower values in general, and DEM\,L71 and \snr{0548}
extend to higher values (see Figure \ref{figure:oiii_hb_sii_ratio_Dec28}).  The [\ion{S}{2}]/H$\alpha$
ratios of the knots range from 0.1 to 1.3. The low values of [\ion{S}{2}]/H$\alpha$ 
ratio are associated with the densest knots whose high densities, $\ge 10^4$
cm$^{-3}$, exceed the critical densities of the \sii $\lambda\lambda$6716, 6731
lines,  as seen in Figure \ref{figure:sii_ha_sii_ratio_Dec28} and elaborated later in Section 5.1.2.

The \nii $\lambda$5755/$\lambda$6584 ratios can be used to diagnose electron temperatures.
The weak \nii $\lambda$5755 line is detected only in knots of N103B and 
DEM\,L71.  N103B has a large number of bright knots and their electron 
temperatures range from 9,000 to 19,000 K.  In DEM\,L71 only the very few 
brightest knots are detected in the \nii\ $\lambda$5755 line and their
temperatures are $\sim$11,000 K.


\begin{figure*}
\epsscale{1.0}
\plotone{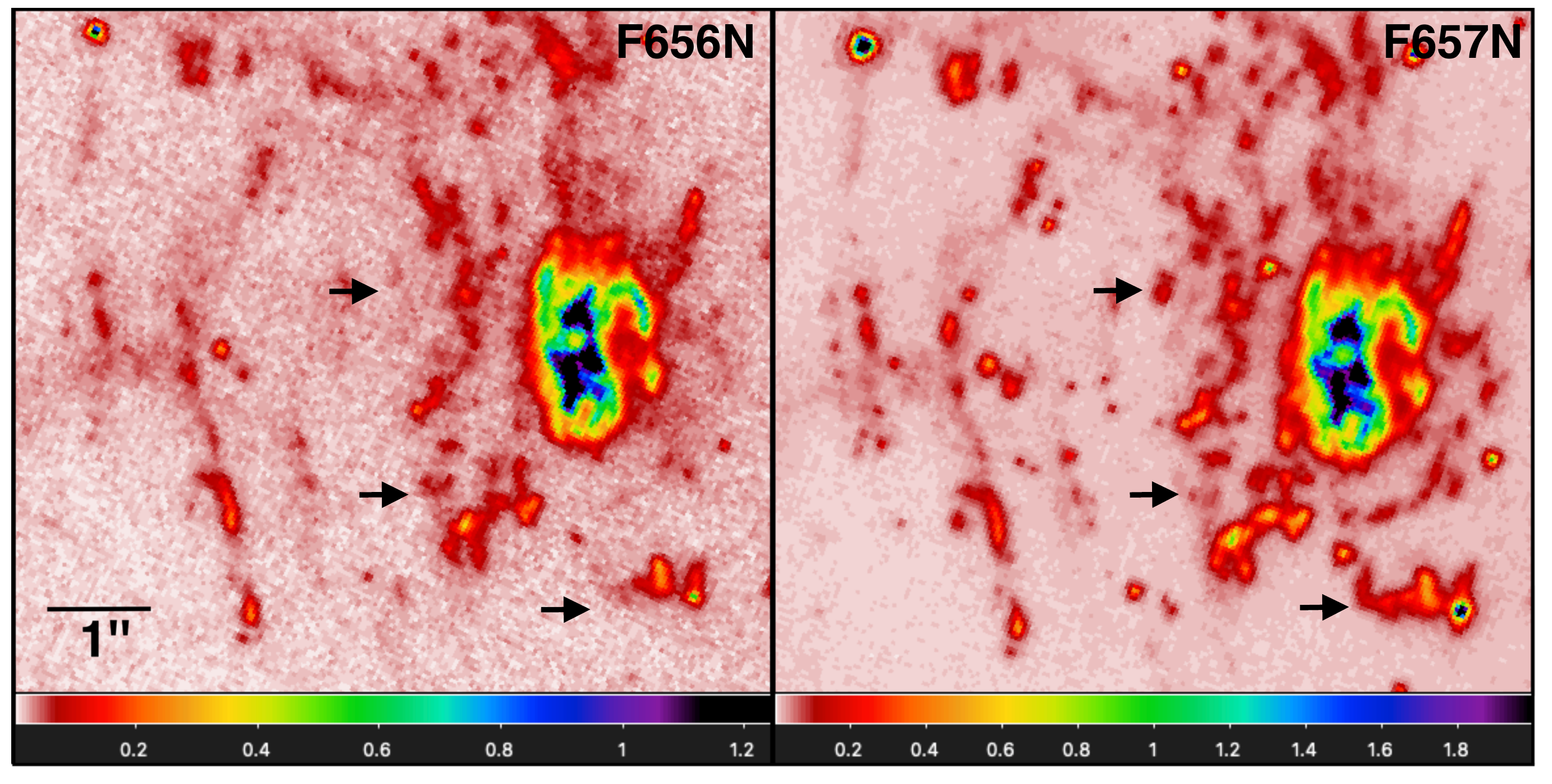}
\caption{ 
The \hst \ha images of the prominent groups of knots within N103B obtained in 2013 (left) and 2017 (right).
Three knots that obviously changed brightnesses are marked by arrows.
}
\label{figure:n103b_hst_ha_3_half_yr_comparison_color2}
\end{figure*}

\subsubsection{Origin and Implication of the Knots}
The knots resolved in HST \ha images of these five Type Ia SNRs are small, dense, and H-rich.
The size, as small as 0.05 pc, and density, as high as $\ge 10,000$ H cm$^{-3}$, are
not characteristic for the ISM. In the HST program 13282, H$\alpha$ images of four 
larger and more evolved Type Ia SNRs are available -- 0454$-$67.2, DEM\,L238, 
DEM\,L249, and DEM\,L316A.  
None of these Type Ia SNRs show small dense knots in their interiors or along their shell rims.  
The HST archive has H$\alpha$ and/or [\ion{O}{3}] images of six core-collapse SNRs in the LMC
-- N49, N63A, and N206 from program 8110, 0540$-$69.3 from programs 6120 and 7340, and
N132D from program 12001, and SN1987A from multiple programs.
Among these, only N63A and N132D have small features that can be compared with the knots
seen in the 5 Type Ia SNRs we have studied. Figure \ref{figure:knots_morphologies} shows that 
the cloudlets in N63A and N132D are qualitatively similar to the knots in the Type Ia SNRs, and they are known to be interacting with molecular clouds \citep{Sano2019,Sano2020}.
However, the rms electron densities in the cloudlets in N63A are 150-700 cm$^{-3}$
\citep{Chu1999} and the electron densities of the knots in N132D are generally 
2000-4000 cm$^{-3}$ \citep{Dopita2018}, much lower than what we see in the knots in Type Ia SNRs with Balmer-Dominated Shells.  SN1987A is transitioning from a SN to an SNR, and only the knots 
in its inner ring have sizes $\sim$0.03 pc and densities 1,000-30,000 
atoms cm$^{-3}$ \citep{Mattila2010}, comparable to the knots we see in Type Ia SNRs.  The rings around SN1987A are known to be of a CSM origin.  We thus consider that the small
knots observed in the Type Ia SNRs with Balmer-Dominated Shells most likely belong to a CSM.

The dense knots in \snr{0519} are not as numerous and wide-spread as those
in N103B, DEM\,L71, and \snr{0548}.  The brightest ones are concentrated in a small 
0\farcs5 $\times$ 3$''$ (0.13 pc $\times$ 3 pc) patch, coincident with the north
patch of [\ion{Fe}{14}] emission.  The east patch of [\ion{Fe}{14}] emission may 
also be associated with dense knots; however, it is superposed on a very bright Balmer
shell rim and its counterparts in [\ion{O}{1}], [\ion{O}{3}], and [\ion{S}{2}] lines 
are all quite weak.
It is likely that the dense knots in \snr{0519} have a different origin from
those in the other three SNRs.  Without further information, such as kinematics 
and elemental abundances, about the knots in \snr{0519}, we cannot confidently
assess how this CSM material was ejected or has evolved.  We will thus discuss 
below mainly knots in the other three Type Ia SNRs. 

Because of the high density in the knots, their recombination time scales are
short.  For example, the recombination timescale is 70 yr for a density of
1000 H cm$^{-3}$ and only 7 yr for a high density of 10,000 H cm$^{-3}$. 
Depending on the propagation of the SNR shocks, different sets of knots should
brighten up and fade as time goes on, much like the knots in SN1987A's inner
ring \citep{Fransson2015}. Indeed, the \emph{HST} H$\alpha$ images
of N103B taken on 2013 July 11 and 2017 Jan 03 show some knots faded and some knots
brightened up, as illustrated in Figure \ref{figure:n103b_hst_ha_3_half_yr_comparison_color2}.

To inter-compare the knots from the Type Ia SNRs, we first examine their 
[\ion{S}{2}]/H$\alpha$ ratio as a function of density in Figure \ref{figure:sii_ha_sii_ratio_Dec28}.
The horizontal axis is \sii $\lambda$6716/$\lambda$6731, which is a diagnostic for electron densities.  
Beside the clear separation of the ISM and CSM densities, it is striking that the knots in 
N103B, DEM\,L71, and \snr{0548} are segregated
in the density distribution.  N103B is the smallest and has the densest knots, while 
\snr{0548} is the largest and has the least dense knots.  This trend is suggestive
of an evolutionary effect: as an SNR evolves and expands to larger size, its CSM
knots go through ionization and shock interactions and become less dense. 
It is also noticeable that the greatest majority of the knots in 
Figure \ref{figure:sii_ha_sii_ratio_Dec28} follow the trend that the 
highest observed  [\ion{S}{2}]/H$\alpha$ ratio for any specific \sii $\lambda$6716/$\lambda$6731 
decreases with decreasing \sii $\lambda$6716/$\lambda$6731, or increasing density.  
This trend is caused by the low critical densities of the \sii lines, 
1585 and 3981 cm$^{-3}$ for $\lambda$6716
and $\lambda$6731, respectively.  While the $\lambda$6716/$\lambda$6731 ratio decreases 
with increasing density, the [\ion{S}{2}]/H$\alpha$ ratio also decreases because of 
increasing collisional de-excitation of the $^2$D levels of S${^+}$.  

The [\ion{O}{3}]/H$\beta$ ratio versus \sii $\lambda$6716/$\lambda$6731 ratio plot in 
Figure \ref{figure:oiii_hb_sii_ratio_Dec28} shows contrasting distributions
among the SNRs. N103B has very few knots with [\ion{O}{3}]/H$\beta$ $>$ 2,
while DEM\,L71 has a significant fraction of knots with [\ion{O}{3}]/H$\beta$
$>$ 2. \snr{0548}, unlike N103B and DEM\,L71, has no knots with 
[\ion{O}{3}]/H$\beta$ $<$ 0.5.  These differences may also be caused by 
their evolutionary status associated with the passage of SNR shocks.
The large spreads of [\ion{O}{3}]/H$\beta$ ratios in 0548$-$70.4 and DEM\,L71 
probably requires incomplete shocks, as suggested by the variability of some of 
the knots mentioned in Section 5.1.2. It is hard to get [\ion{O}{3}]/H$\beta$ 
above 2 or 3 except by having an incomplete recombination zone. The very 
low [\ion{O}{3}]/H$\beta$ ratios in some of the N103B knots (below about 0.2) 
suggest very slow shocks, less than 100 kms$^{-1}$.

 It is tempting to conclude that N103B, DEM\,L71, and \snr{0548} 
all contain dense CSM and have single-degenerate SN progenitors, 
although surviving companions of the progenitors have not been successfully
identified \citep{Li2019}.  The absence of an abundant CSM in \snr{0509} and \snr{0519}
has been used to argue that their SN progenitors were likely of double-degenerate
origin, especially since the young \snr{0509} has no viable stellar candidate
for its SN progenitor's companion near the center of the SNR \citep{Schaefer2012, Litke2017}.
However, the discovery of the knots in \snr{0519} raises question about their origin,
which may not be entirely clear at present.

\subsection{Collisionless Shocks of the Balmer Shells}

Both \snr{0509} and \snr{0519} have been well-studied: their shock speeds have been measured from HST proper motions by \citet{Hovey2018}, and their ambient preshock densities have been derived from dynamical models by \citet{Seitenzahl2019}.
We will focus on the case of \snr{0519} because its [\ion{O}{3}]
emission is detected and the [\ion{O}{3}]/H$\beta$ ratio has a high S/N. 
We adopt shock speeds of 1300--2500 km s$^{-1}$ and a preshock density of 
1.5 $\rm amu\, cm^{-3}$ from the above references.
We assume the hydrogen to be 50\% neutral because if it were highly ionized the Balmer 
filaments would be correspondingly faint, and because helium atoms swept up by the shock 
produce enough photons to ionize at least 30\% of the hydrogen.  
We concentrate on [\ion{O}{3}], but generally similar considerations would apply to [\ion{S}{2}].

Balmer line filaments show essentially pure Balmer line spectra in the optical range.  
The hot gas behind the shock efficiently excites neutral H atoms that pass through 
the shock before they can be ionized.  
Thus the profiles of the Balmer lines show two components: a broad component whose 
width is related to the postshock proton temperature and a narrow component whose 
width reflects the preshock temperature.
For postshock temperatures of 10$^6$ K or more, each neutral hydrogen atom produces 
about 0.25 H$\alpha$ and about 0.05 H$\beta$ photons on average before it is ionized.  
By comparison, there are several thousand times fewer oxygen atoms than hydrogen, 
so the forbidden lines that dominate most SNR shock wave spectra are expected to be faint.   

 Most shocks in the ISM except for C-shocks \citep{Draine1983} are collisionless, 
 meaning that the shock transition is governed by magnetic fields and plasma turbulence.   
 The shock thickness is determined by the proton gyroradius or the ion skin depth, 
 which are far smaller than the particle mean free path.  Because the bulk velocities 
 of particles entering the shock are randomized by plasma processes rather than 
 collisions, they do not reach thermal equilibrium.  Instead of Maxwellian velocity 
 distributions, both electrons and ions can show high energy tails as a result of 
 diffusive shock acceleration \citep{Blandford1987}, and different particle species 
 can have  different temperatures.  In shocks faster than 1000 km s$^{-1}$, electron 
 temperatures $T_{\rm e}$ are only a few percent as high as the proton temperatures 
 $T_{\rm p}$ \citep{Ghavamian2001, Ghavamian2013}, and ion temperatures $T_{\rm i}$ 
 tend to be proportional to ion mass $m_{\rm i}$, 
 $T_{\rm i} \sim (m_{\rm i}/m_{\rm p})\,T_{\rm p}$ \citep{Korreck2004, Raymond2017}.  
 The collisionless nature of these shocks also 
 explains the two-component Balmer line profiles.  Some neutrals pass though the 
 shock unaffected, so when they are excited they produce emission with the preshock 
 velocity profile.  Others experience charge transfer with a postshock proton, and 
 when they are excited they produce a correspondingly broad profile.

There are two possible origins for the [\ion{O}{3}] emission.  It could be produced 
in a narrow ionization zone just behind the shock in the same manner as the Balmer 
lines, or it could arise from a shock precursor.  If the [\ion{O}{3}] were produced 
in the shocked gas, the line would be very broad.  Since there is little thermal 
equilibration in such fast shocks, the FWHM of the emission from 0519--69.0 would be similar to the shock speeds of 1300 \kms at the E limb and 2500 \kms at the S limb, or 25 to 60 \AA. That would 
mean that little of the emission would be within the  8 \AA\ band of the 
[\ion{O}{3}] image, and the emission in the off-band image that is subtracted 
would nearly cancel out what remains.  

Given an LMC oxygen to hydrogen ratio  of 0.0002 to 0.0003 \citep{Russell1992} and an emission rate 
of 0.05 H$\beta$ photons per hydrogen atom, the observed [\ion{O}{3}] to H$\beta$  
ratio of 0.011 would require about 0.7 [\ion{O}{3}] photon per O atom.  At very 
high temperatures, the number of photons is given by the ratio of excitation 
rate to ionization rate.  We take excitation and ionization rates by electrons 
from CHIANTI version 8 \citep{Dere1997, DelZanna2015} and we assume that the 
excitation rate by protons is about equal to that by electrons at $T_{\rm e} = 
(m_{\rm e}/m_{\rm p})\,T_{\rm i}$  to find that each O atom should produce 
0.3 to 0.5 [\ion{O}{3}] photons. It therefore seems that the postshock region 
could produce almost the observed [\ion{O}{3}] intensity if the intensity were not 
spread over a band much wider than that used to generate  Figure~\ref{figure:line_imgs_0519}. 
We conclude that the [\ion{O}{3}]\ emission in 0519--69.0 cannot come from the postshock gas.  

The alternative emission region is a shock precursor.  Three kinds of precursor 
can be present;  a photoionization precursor, a cosmic ray precursor associated 
with diffusive shock acceleration \citep{Blandford1987, Boulares1988},  or a 
precursor produced by broad component neutrals overtaking the shock and depositing 
their energy upstream \citep{Hester1994, Morlino2012}.  Precursors have been 
inferred from \ha narrow components that are broader than would be consistent 
with a significant neutral fraction in equilibrium, indicating that the gas is 
heated in a narrow precursor just ahead of the shock \citep{Hester1994}, and 
they can be seen as faint emission ahead of the main shock in Tycho’s SNR 
\citep{Lee2010}.  The proton kinetic temperatures indicate narrow component widths 
ranging from about $3 \times 10^4$ to $10^6$ K \citep{Sollerman2003, Medina2014, 
Knezevic2017}, but the electron temperatures are poorly constrained.  
The precursor thickness ranges from $0.3 \times 10^{16}$ cm  to 
$3 \times 10^{16}$ cm \citep{Katsuda2016, Lee2010}.  
Photoionization precursors associated with Balmer line filaments have been 
reported for Tycho’s SNR \citep{Ghavamian2000} and the Cygnus Loop \citep{Medina2014}.
The photoionization is dominated by \ion{He}{1} and \ion{He}{2} photons near 21 and 
40 eV, respectively, and they heat the gas to around 17,000 K.

The existence of the Balmer line filaments implies a substantial neutral hydrogen fraction in the
upstream gas. \ion{O}{3}\ does not coexist with neutral H because of the very rapid charge transfer
process H$^0$ $+$ O$^{++}$ $\rightarrow$ H$^+$ $+$ O$^+$, which has a rate coefficient of 1.0$\times$10$^{-9}$ cm$^{3}$ s$^{-1}$ \citep{Kingdon1996}. Therefore, oxygen must be ionized to \ion{O}{3}\ in the precursor. It cannot be collisionally
ionized, because hydrogen would be more rapidly ionized, and no neutral atoms would reach
the shock to produce the \ha broad component. However, \ion{O}{3}\ can be produced by
photoionization by \ion{He}{2}\ $\lambda$304 photons, because the photoionization cross 
section of O$^+$ at a
photon energy of 40 eV is 30 times larger than that of H$^0$ \citep{Reilman1986}. 
For a density of 
1.5 cm$^{-3}$ and 80\% ionization, the lengths for absorption by a mixture of H and He  
correspond to 1\farcs5 and 7$''$ for \ion{He}{1} and \ion{He}{2} photons, respectively.
The latter is similar to the precursors seen in the E and S limbs of 
Figure~\ref{figure:line_imgs_0519}, 
while the former is compatible with the thickness of the [\ion{O}{3}] emitting region, 
though that may be dominated by projection effects
rather than the physical thickness. 
The flux of \ion{He}{2}\ $\lambda$304 photons from the shock is
proportional to the number of He atoms swept up per second, and therefore to the shock
speed. Only shocks faster than about 1000 \kms\ produce enough photons to dominate
over the charge transfer rate, which explains why [\ion{O}{3}]\ is detected in the fastest shocks.


\begin{figure}
\epsscale{2.3}
\plottwo{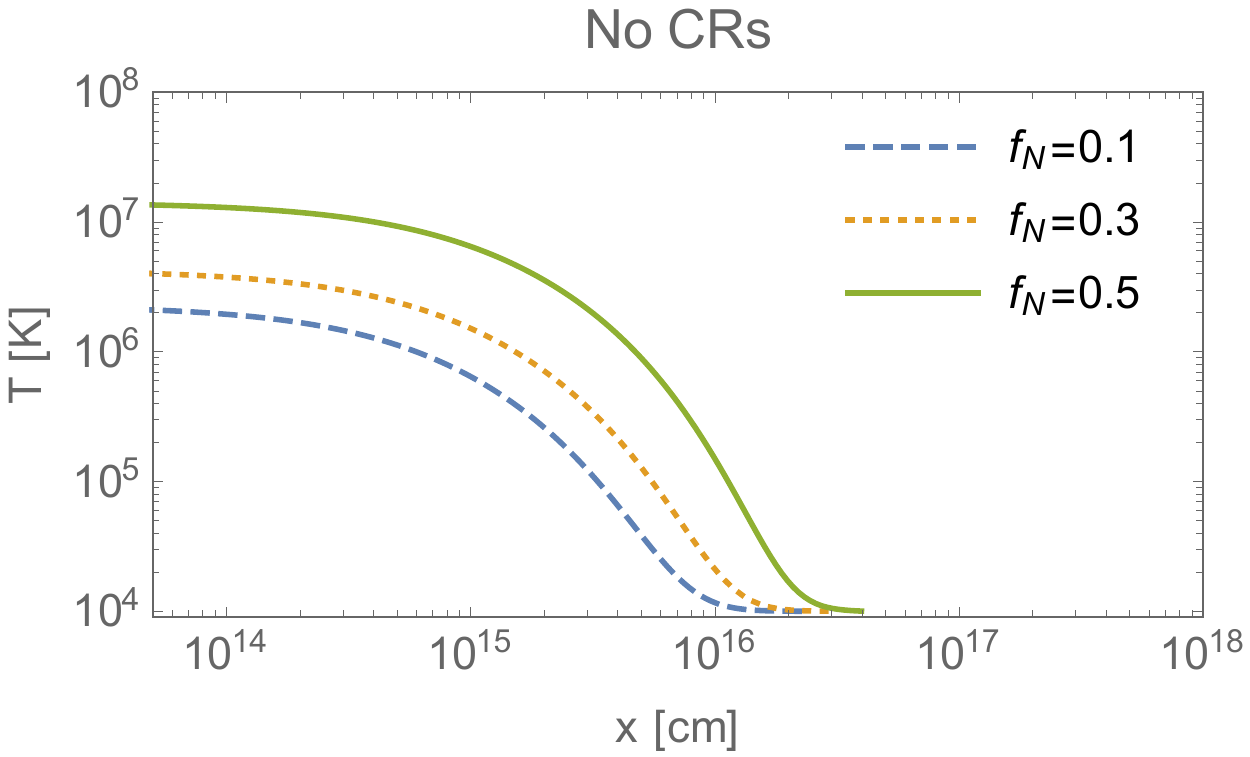}{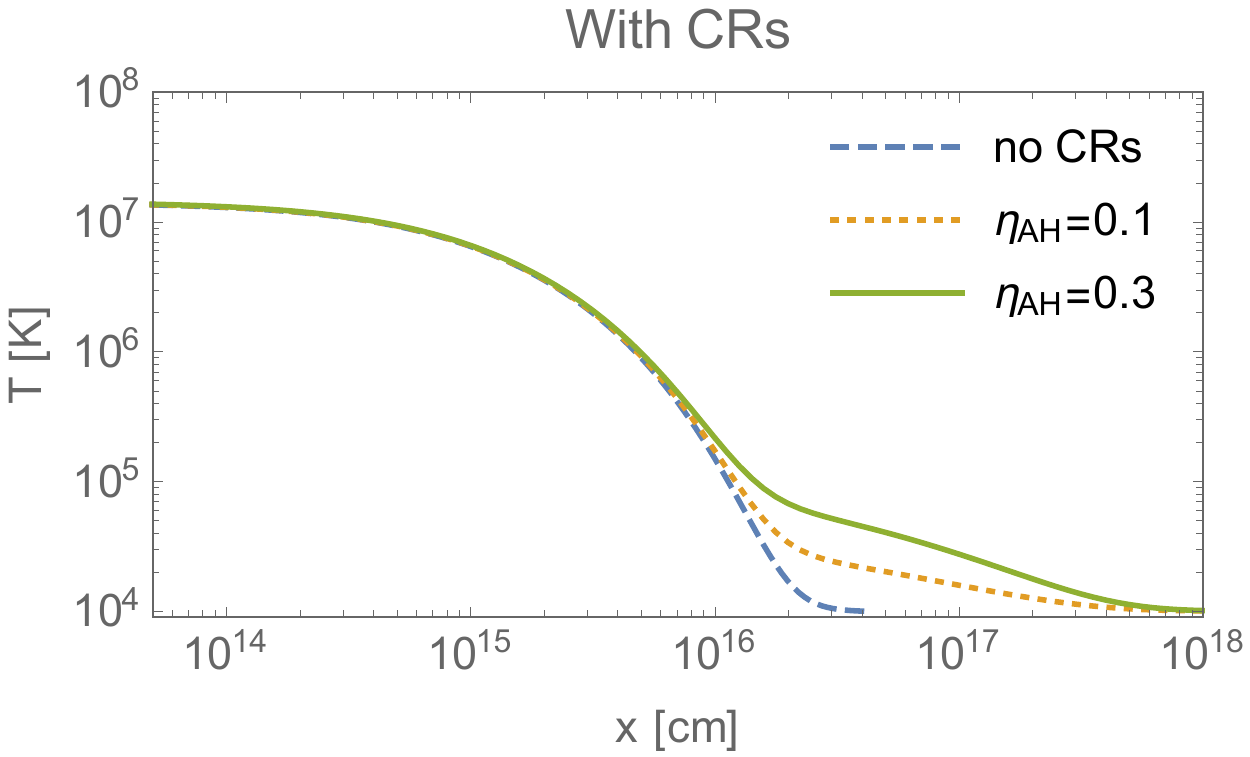}
\caption{
Temperature profile of protons inside the precursor with (bottom panel) and without (upper
panel) cosmic rays. For all cases the assumed shock velocity is V$_{sh}$ = 2500 \kms\ and the ambient density is
 1.5 cm$^{-3}$.
Upper panel: No cosmic rays. The three lines correspond to three different neutral fraction (10, 30 and
50\%).
Bottom panel: temperature profile assuming cosmic rays accelerated with an efficiency of $\sim$5\% and a
neutral fraction of 50\%. The two lines show the effect of changing the wave damping efficiency
from 0.1 to 0.3. 
}
\label{figure:T1_CRs}
\end{figure}

To evaluate the different roles of neutral and cosmic ray precursors, we calculate the temperature
ahead of the shock with and without cosmic rays acceleration. We use the model developed by \citet{Morlino2013} which allows us to calculate the shock structure in the presence of both neutral H and
cosmic rays. For all cases analyzed we assumed a shock velocity of 2500 \kms\ and an ambient density of 1.5
cm$^{-3}$. Figure \ref{figure:T1_CRs} shows the resulting temperature profile of H ions ahead of the shock. The top
panel shows the cases without cosmic rays where we change the neutral fraction from 10\% up to 50\%.
The precursor temperature increases with increasing neutral fraction as a consequence of the
larger number of neutrals returning from the downstream. For a neutral fraction $=$ 50\% the
precursor temperature is $\sim$ 10$^7$ K hence it would be enough to efficiently excite \ion{O}{3}. 
Nevertheless the precursor thickness is only $\lesssim$ 10$^{16}$ cm (corresponding to
0\farcs013), roughly an order of magnitude smaller than the excitation lengthscale for O. Moreover, the very efficient heating by
backstreaming neutrals raises the temperature to about $10^7$ K, and at that temperature
O III is ionized more rapidly than it is excited.

The bottom panel of the same figure shows, instead, the impact of including cosmic ray acceleration.
We assume that the acceleration efficiency is only 5\%, in order to be compatible with the upper
limit of $\sim$ 7\% found by \cite{Hovey2018} and that the maximum energy of accelerated protons
reaches $E_{\max}= 1$ TeV. Such energies can be reached if the diffusion coefficient is Bohm-like in a $\sim$ $1 \mu$ G field. Notice that the maximum extent of the cosmic ray precursor corresponds to
the propagation lengthscale at $E_{\rm max}$, which is $l_{\rm d} = D_{\rm Bohm}(E_{\max}, 1
\mu \rm G)/V_{\rm sh} \simeq 4 \times 10^{17}\, {\rm cm}$.
We show two cases where the neutral fraction is fixed to 50\% but the wave damping rate
changes from $\eta_{AH}$=0.1 to 0.3. The parameter $\eta_{AH}$ is the fraction of magnetic
waves's energy (excited by cosmic rays) which is damped to thermal energy of the plasma and
determines the final precursor temperature.
It is worth noting that similar temperature profiles can be obtained for different combinations
of acceleration efficiency, $\eta_{AH}$, $E_{\max}$ and magnetic field. But, what is important
in this context is that, for reasonable values of the parameters (even for relatively small
acceleration efficiency), the precursor temperature starts increasing already at a distance larger
than 10$^{17}$ cm, making possible the effective collisional excitation of \ion{O}{3}\ lines.

We conclude that photoionization is responsible for the presence of O$^{++}$ in the gas upstream of
the shock, but heating by the cosmic ray precursor, and to a lesser extent by the return neutral
precursor, increases the temperature and doubles the excitation rate of the [\ion{O}{3}]\ transitions.
For the excitation rate at temperatures of 20,000 to 40,000 K and a preshock density of 1.5 cm$^{-3}$, the
observed [\ion{O}{3}]\ flux in 0519-69.0 indicates an \ion{O}{3}\ ionization fraction of around 5\% to 20\%. Thus all
three types or precursor contribute to the observed [\ion{O}{3}]\ brightness.

\section{Summary} \label{sec:Summary}  

Five Type Ia SNRs with Balmer-Dominated Shells are known in the LMC:
\snr{0509}, \snr{0519}, N103B, DEM\,L71, and \snr{0548}.
We have been using HST images and VLT MUSE observations
of these SNRs to search for surviving companions of their SN 
progenitors \citep{Litke2017,Li2017,Li2019}.  In the course of 
this work, we find prevalent forbidden emission from these
Type Ia SNRs.  Three types of forbidden line emission
are detected: bright [\ion{O}{3}], [\ion{O}{1}], [\ion{S}{2}],
etc.\ and faint [\ion{Fe}{14}] emission from shocked dense 
knots interior to the SNR shells, [\ion{Fe}{14}] and similar 
high Fe ion lines from reverse shocks into the SN ejecta, 
and faint [\ion{O}{3}] and other low-ionization line 
emission from the Balmer shells.  

Small dense knots are detected in all except \snr{0509}.  
MUSE spectra of the knots show bright forbidden line
emission from [\ion{O}{3}], [\ion{N}{2}], [\ion{S}{2}], etc.
Electron densities determined from the [\ion{S}{2}] 
$\lambda$6716/$\lambda$6731 doublet ratio ranges from 
a few hundred to $\ge 10^4$ cm$^{-3}$.  The knots can be
slightly elongated or rod-like with major-to-minor
axis ratios greater than 10. As the densities exceed 
the critical densities of the [\ion{S}{2}] $\lambda\lambda$6716, 6731
lines, the [\ion{S}{2}]/H$\alpha$ ratio decreases at high densities
and reaches as low as 0.1.  The high density and small size of the
knots are not characteristics of ISM; thus, the knots must have a
CSM origin.  

The recombination time scale for the dense knots are 
below 10 years, and indeed brightness variations in knots can be 
seen in the HST images of N103B from two epochs separated 
by 3.5 yr.  Physical properties of the dense knots vary among the
five Type Ia SNRs we studied; the variations appear to be correlated with 
the SNR ages (in the second column of Table \ref{table:ratio}).  
\snr{0509} has no knots at all, \snr{0519} has a small
patch of knots and possibly another faint patch, N103B has the most 
prominent knots, DEM\,L71 displays knots near the shell rim, and 
\snr{0548} has knots already shredded and disperse.  It is conceivable 
that as time goes on, the CSM knots will be shocked to light up, then 
recombine and dissipate.  The presence of CSM in Type Ia SNRs could 
be more prevalent than we previously thought.

The faint [\ion{O}{3}] line emission from the Balmer shell is 
detected in VLT MUSE observations of SNRs \snr{0519}, N103B,
and DEM\,L71. The ATT WiFeS
observations of \snr{0548} have too limited spatial coverage to
determine whether this SNR also has [\ion{O}{3}] emission from 
its Balmer shell.  We focus on the case of \snr{0519} because its
[\ion{O}{3}] emission is well-measured and its shock velocity
and ambient ISM density have been studied in detail and reported
by \citet{Hovey2018} and \citet{Seitenzahl2019}.  We exclude the 
postshock origin of the [\ion{O}{3}] emission because its FWHM 
is not compatible with the broad component of the Balmer lines.  
For the preshock origin, we considered three possibilities:
photoionization precursor, cosmic ray precursor, and neutral 
precursor.  With considerations of the [\ion{O}{3}]/H$\beta$ ratio 
and thickness of the [\ion{O}{3}] shell, 
we conclude that the [\ion{O}{3}] emission arises from oxygen that has been photoionized by [\ion{He}{2}] $\lambda$304 photons and is then collisionally excited in a shock precursor heated mainly by cosmic rays.
A more detailed quantitative analysis of the nebular lines in these Type Ia SNRs will be
carried out and reported in a future paper.

\acknowledgments
Y.-H.C. and C.-J.L. are
supported by Taiwanese Ministry of Science and Technology
grant MOST 108-2811-M-001-587, 109-2112-M-001-040, and 109-2811-M-001-545.

\emph{Software:} SAOImage DS9 \citep{Joye2003}, QFitsView \citep{Ott2012}, astropy \citep{Astropy2018}, matplotlib \citep{Hunter2007}, numpy \citep{vanderWalt2011, Harris2020}, scipy \citep{Virtanen2020}

\appendix


\begin{figure*}
\epsscale{1.1}
\hspace{-1.cm}
\plotone{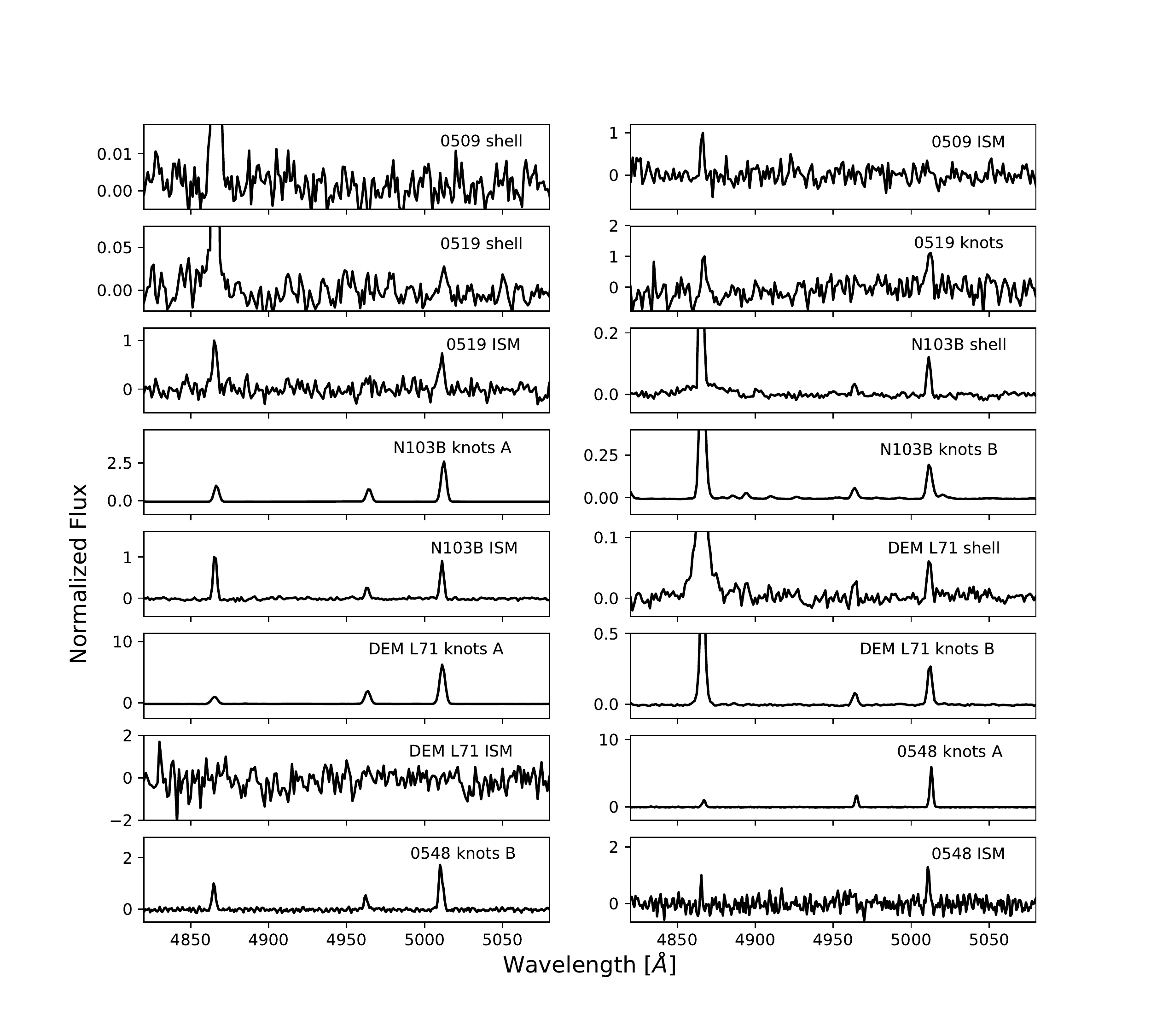}
\caption{  
The spectra obtained from VLT MUSE and ATT WiFeS observations showing the representative \oiiitohb ratios in and around SNRs with Balmer-dominated shells. In each panel, all emission lines fluxes are normalized to the \hb line. Knots A and B are examples with contrasting \oiiitohb ratios.
}
\label{figure:spec_oiii_hb_ratio}
\end{figure*}


\begin{figure*}
\epsscale{1.1}
\hspace{-1.cm}
\plotone{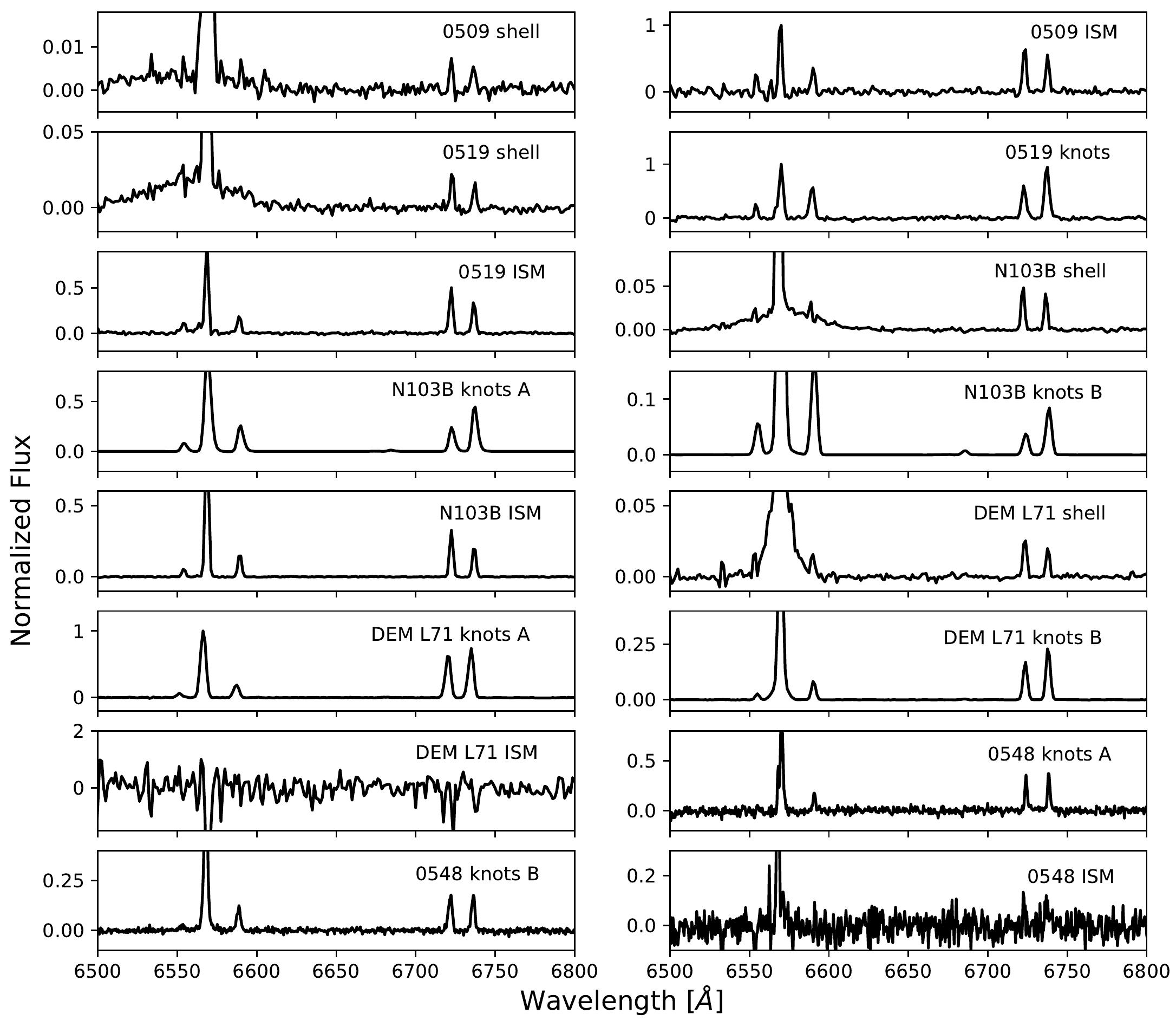}
\caption{
The spectra obtained from VLT MUSE and ATT WiFeS observations showing the representative \siitoha ratios in and around SNRs with Balmer-dominated shells. In each panel, all emission lines fluxes are normalized to the \ha line. Knots A and B are examples with contrasting \siitoha ratios.
}
\label{figure:spec_sii_ha_ratio}
\end{figure*}

\end{document}